\overfullrule=0pt
\input harvmac
\input amssym.tex

\def\ad{{\dot a}}
\def\bd{{\dot b}}
\def\cd{{\dot c}}
\def\l{{\lambda}}
\def\b{{b}}
\def\a{{a}}

\def\k{{\kappa}}

\def\d{{\delta}}

\def\e{{\epsilon}}
\def\s{{\sigma}}

\def\half{{1\over 2}}
\def\p{{\partial}}
\def\pb{{\bar\partial}}
\def\t{{\theta}}

\def\bar{\overline}
\def\hat{\widehat}

\Title{\vbox{\hbox{IFT-P.019/2004 }}}
{\vbox{
\centerline{\bf Conformal Supergravity in Twistor-String Theory}}}
\bigskip\centerline{Nathan Berkovits$^*$
}
\bigskip
\centerline{\it Instituto de F\'\i sica Te\'orica, Universidade
Estadual Paulista} \centerline{\it Rua Pamplona 145, 01405-900,
S\~ao Paulo, SP, Brasil} \smallskip\centerline{and
}\smallskip\centerline{Edward Witten$^{\#}$}
\bigskip
\centerline{\it School of Natural Sciences, Institute for Advanced
Study}
\centerline{\it Princeton NJ 08540, USA}
 \vskip .3in
\noindent
 Conformal supergravity arises in presently known
formulations of twistor-string theory either via closed strings or
via gauge-singlet open strings.  We explore this sector of
twistor-string theory, relating the relevant string modes to the
particles and fields of conformal supergravity. We use the
twistor-string theory to compute some tree level scattering
amplitudes with supergravitons. Since the supergravitons interact
with the same coupling constant as the Yang-Mills fields,
conformal supergravity states will contribute to loop amplitudes
of Yang-Mills gluons in these theories.  Those loop amplitudes
will therefore not coincide with the loop amplitudes of pure super
Yang-Mills theory.

 \vskip 1cm

* {e-mail: nberkovi@ift.unesp.br},   \# e-mail: witten@ias.edu

\Date{June 2004}

\newsec{Introduction}

The most commonly studied supergravity theories are ``Einstein''
supergravity theories, in which the gravitational part of the
action, in $n$ dimensions, is $\int d^nx \sqrt{g} R$, with $R$ the
Ricci scalar. These theories are not special to four dimensions;
they exist up to eleven dimensions \ref\cjs{E. Cremmer, B. Julia,
and J. Scherk, ``Supergravity Theory In Eleven Dimensions,'' Phys.
Lett. {\bf B79} (1978) 231.}.

In four dimensions, there also exist conformally invariant
supergravity theories (see \ref\confgrav{E.S. Fradkin and A.A. Tseytlin,
``Conformal Supergravity,'' Phys. Rept. {\bf 119} (1985) 233.}
for a review),
%NB added reference
in which the gravitational part of the
action is $\int d^4x \sqrt g W^2$, with $W$ the Weyl tensor.  Such
theories are special to four dimensions, in the following sense.
In $n$ dimensions, the conformally invariant expression would be
$\int d^n x \sqrt g W^{n/2}$ if one does not consider terms involving
derivatives of $W$.\foot{In six dimensions, as pointed out to us by K. 
Skenderis, one can write
a conformally invariant action of the form $\int d^6x \sqrt{g}( W \square W+ 
...)$, which leads to Euler-Lagrange equations that are of sixth order, as 
opposed to the fourth order equations coming from the usual conformally 
invariant action $\int d^n x \sqrt{g}W^{n/2}$.} Precisely for $n=4$, this
expression becomes quadratic in $W$ and gives a nondegenerate
kinetic energy for the gravitational field.

This simple observation suggests that four-dimensional conformal
supergravity theories might be relevant to the real world --
perhaps with the aid of some mechanism of spontaneous breaking of
conformal invariance. However, they in fact are generally
considered to be an unsuitable starting point for describing
nature, because they lead to fourth order differential equations
for the fluctuations of the metric, and thus to a lack of
unitarity. We have no reason to question these beliefs and we will
later describe some facts that illustrate them.

\nref\witten{E. Witten, ``Perturbative Gauge Theory As A String
Theory In Twistor Space,''  hep-th/0312171. }%
\nref\berk{N. Berkovits, ``An Alternative String Theory In Twistor Space For
${\cal N}=4$ Super Yang-Mills,'' hep-th/0402045.}%

 The usual string theories are not
conformally invariant in the target space, and give at low
energies Einstein supergravity rather than conformal supergravity.
Twistor-string theory \refs{\witten,\berk} is clearly different;
superconformal invariance in spacetime is built in, and thus any
form of supergravity that emerges will be conformal supergravity.

The first indication that conformal supergravity arises in
twistor-string theory was presented in section 5.1 of  \witten,
where it was shown that tree level gluon scattering amplitudes
contain, in additional to the single-trace Yang-Mills scattering
amplitudes, multi-trace terms that reflect the exchange of
conformal supergravitons.  At tree level, it is possible to
recover the pure Yang-Mills scattering by extracting the
single-trace amplitudes. However, at the loop level, diagrams that
include conformal supergravitons can generate single-trace
interactions, so the presence of conformal supergravity apparently
means that it will be difficult with presently known forms of
twistor-string theory to compute Yang-Mills or super Yang-Mills
amplitudes beyond tree level.
 In twistor-string theory, conformal supergravitons have the same coupling constant
 as gauge bosons, so it is not possible to remove the conformal supergravity contributions
 to scattering amplitudes by going to weak coupling.\foot{To be more precise,
 in one approach to twistor-string theory \witten\ this is true as stated,
 while in the other approach \berk, as we discuss in section 7, the ratio of
 the gravitational and gauge coupling constants is proportional to $k$, the
 level of the current algebra.  Unitarity requires $k\geq 1$, but if it makes sense
 to relax unitarity, one could conceivably decouple conformal gravity in the limit
 $k\to 0$.} Since Yang-Mills
theory makes sense without conformal supergravity, it is plausible
that a version of twistor-string theory might exist that does not
generate conformal supergravity and would be useful for computing
Yang-Mills loop amplitudes.  It would be highly desireable to find
such a theory, though at the moment it is not clear how to do so.

The occurrence of conformal supergravity in twistor-string theory
seems interesting and unusual enough to be worthy of study, even
though conformal supergravity appears not to be physically
sensible. In this paper, we consider two alternative versions of
twistor-string theory based on the $B$-model of $\Bbb{CP}^{3|4}$
\witten\ and based on a certain construction involving open strings \berk.
Since gravity usually arises in the closed string sector, the
appearance of conformal supergravity in the open version of twistor-string
theory is somewhat unexpected.
%NB added above sentence
(It would also be interesting to reconsider some of the issues using a
more recent alternative twistor-string proposal \ref\siegel{W.
Siegel, ``Untwisting The Twistor Superstring,'' hep-th/0404255.}, as
well as the twistor-string proposal of \ref\vafa{
A. Neitzke and C. Vafa, ``N=2 Strings and the Twistorial
Calabi-Yau,'', hep-th/0402128\semi M. Aganagic and C. Vafa,
``Mirror Symmetry and Supermanifolds,'' hep-th/0403192.}.)
%NB added reference to Vafa papers and started new paragraph

We consider various issues involving conformal supergravity in the
two models, showing how rather similar results arise from
different origins. In sections 2 and 3, we analyze the spectrum of
the theory first in twistor space and then in Minkowski spacetime.
In section 4, we discuss the linearized spectrum of conformal
supergravity and compare to the twistor-string results. In section
5, we use the twistor-string theory to  compute some tree
amplitudes including conformal supergravitons.   In section 6, we
discuss some properties of the nonlinear conformal supergravity
action. Finally, in section 7, we analyze anomalies that
apparently lead to restrictions on the gauge group. Our discussion
in section 7 is somewhat inconclusive.

\newsec{Vertex Operators}

In this section, we will describe the twistor-string vertex
operators. First we consider the ``open string'' version of
twistor-string theory \berk, and then we consider the $B$-model of
$\Bbb{CP}^{3|4}$ \witten.

\subsec{``Open String'' Version}

In the open version of the twistor string, the worldsheet action
is \eqn\actions{ S = \int d^2 z ( Y_I \pb_A Z^I + \bar Y{}_I
\partial_A\bar Z{}^I + S_C ).  } Here, assuming a Euclidean signature
worldsheet,\foot{ With a Lorentz signature worldsheet, one can
alternatively  treat $Z^I$ as homogeneous coordinates of
$\Bbb{RP}^{3|4}$, and $\bar Z^I$ as homogeneous coordinates of a
second $\Bbb{RP}^{3|4}$.   The gauge group is $GL(1,\Bbb{R})\times
GL(1,\Bbb{R})$, with separate real scalings of $Z$ and $\bar Z$.
The open string boundary condition is the same, $Z^I=\bar Z^I$,
and the open string vertex operators are accordingly also the
same.} $Z^I$ are homogeneous coordinates of $\Bbb{CP}^{3|4}$,
$\bar Z^I$ are their complex conjugates, and $Y$ and $\bar Y$ are
variables conjugate to $Z$ and $\bar Z$, respectively. $Z$ and
$\bar Z$ have conformal dimensions $(0,0)$, while $Y$ and $\bar Y$
have dimensions $(1,0)$ and $(0,1)$.  $A$ is a worldsheet gauge
field that gauges the $GL(1,\Bbb{C})$ symmetry $Z^I\to t Z^I$,
$Y_I\to t^{-1} Y_I$. $S_C$ is the action for an additional system
with $c=28$. We assume that this includes a current algebra of
some group $G$, which will become a gauge group in spacetime, and
we refer to the variables in $S_C$ as current algebra variables.
For the open string, \eqn\hofot{Z^I=\bar Z{}^I,~Y_I= \bar Y{}_I}
 on
the boundary.  On the boundary, the gauge group $GL(1,\Bbb{C})$
(or $GL(1,\Bbb{R})^2$ in the case of a Lorentz signature
worldsheet, as remarked in the footnote) is broken to
$GL(1,\Bbb{R})$, the group of real scalings of $Z$ and $\bar Z$
that preserve the boundary condition.

Physical states are described by dimension one  fields  or vertex
operators that are neutral under $GL(1)$ and moreover are  primary
fields with respect to the Virasoro and GL(1) generators.  These
generators are \eqn\vir{T = Y_I \p Z^I + T_C, \quad J = Y_I Z^I,}
where $T_C$ is the $c=28$ stress tensor for the current algebra.
We consider mainly open string vertex operators.  The boundary
condition \hofot\ means that open string vertex operators can be
expressed in terms of $Z$ and $Y$ and the current algebra
variables, and not $\bar Z$ and $\bar Y$; moreover, by virtue of
the boundary condition, $Z$ and $Y$ are real on the boundary. (For
a Lorentz signature worldsheet, $Z$ and $Y$ are real even away
from the boundary, as noted in the footnote.)

The most obvious primary fields are the dimension zero fields
$\phi(Z^I)$, with $\phi$ being any function that is invariant
under $GL(1,\Bbb{R})$ scalings of $Z^I$ (in other words, $\phi$ is
invariant under $Z\to t Z$ for real $t$); equivalently, $\phi$ is
any function on $\Bbb{RP}^{3|4}$. By multiplying such a field by
any of the currents $j_r$, $r=1,\dots,{\rm dim}\,G$ of the current
algebra, we can construct Yang-Mills vertex operators, which of
course should have dimension 1:
 \eqn\ymv{V_\phi = j_r \phi^r(Z).}
These vertex operators were discussed in \berk\ in reproducing
some of the results of \witten\ and \ref\nair{V. P. Nair, ``A
Current Algebra For Some Gauge Theory Amplitudes,'' Phys. Lett.
{\bf B214} (1988) 215.}.

With equal ease, one can construct the vertex operators that turn
out to describe conformal supergravity.  An expression linear in
either $Y$ or $\partial Z$ has dimension 1. So the following
operators have dimension 1: \eqn\vertex{V_f = Y_I f^I(Z),\quad V_g
= g_I(Z) \p Z^I.} These are in addition $GL(1)$-invariant if $f^I$
carries  GL(1) charge $1$ (that is, under $Z\to t Z$, it scales as
$f\to tf$) and $g_I$ carries  GL(1) charge $-1$ (it scales as
$g\to t^{-1}g$). To be primary fields with respect to $J$ and $T$,
$f^I$ and $g_I$ must satisfy \eqn\condit{\p_I f^I=0,\quad Z^I
g_I=0.} Furthermore, $f^I$ and $g_I$ have the gauge invariances
\eqn\gauge{\d f^I = Z^I \Lambda,\quad \d g_I = \p_I\chi} since
$Y_I Z^I \Lambda = J_{-1} \Lambda$ and $\p Z^I \p_I \chi = T_{-1}
\chi$.

These conditions have a simple interpretation.  Since $f^I$ has
charge 1, the expression
\eqn\honon{\Upsilon=f^I{\partial\over\partial Z^I}} is invariant
under scaling.  The equivalence relation $\d f^I=Z^I\Lambda$ means
that $\Upsilon$ can be interpreted as a vector field on
$\Bbb{RP}^{3|4}$ (and not on the ambient $\Bbb{R}^{4|4}$).  The
constraint $\p_I f^I=0$ means that it is a {\it volume-preserving}
vector field.

In the case of $g$, the natural expression is the one-form
\eqn\gonon{\Theta=g_I\,dZ^I.} The constraint $g_IZ^I=0$ means that
$\Theta$ is well-defined as a one-form on $\Bbb{RP}^{3|4}$ (and
not just on $\Bbb{R}^{4|4}$).  The gauge equivalence $\d
g_I=\p_I\chi$ means that $g$ can be regarded as an abelian gauge
field on $\Bbb{RP}^{3|4}$, not just a one-form.  Of course,  like
the functions $\phi^r $ in \ymv\ that describe gauge fields in
spacetime or the volume-preserving vector field $f^I$, the abelian
gauge field $g_J$ is not constrained to obey any equation of
motion.

\subsec{The $B$-Model Of $\Bbb{CP}^{3|4}$}

Now we consider the analogous issues in the other version of
twistor-string theory -- the $B$-model of $\Bbb{CP}^{3|4}$.

We denote the complex homogeneous coordinates of $\Bbb{CP}^{3|4}$
as $Z^I=(\lambda^a,\mu^{\dot a},\psi^A)$, where $\lambda^a$ and
$\mu^{\dot a}$ are bosonic spinors of opposite helicity, and
$\psi^A$ are fermions.  $\Bbb{CP}'^{3|4}$ denotes the region in
$\Bbb{CP}^{3|4}$ in which the $\lambda^a$ are not both zero.  The
usual twistor space wavefunctions (for examples, wavefunctions
corresponding to plane waves in Minkowski spacetime) are regular
on $\Bbb{CP}'^{3|4}$.

The analysis in \witten\ centered on the open string states of the
$B$-model that are related to gauge fields in spacetime.  As
suggested by the Penrose transform \ref\penrose{R. Penrose,
``Twistor Quantization And Curved Spacetime,'' Int. J. Theor.
Phys. {\bf 1} (1968) 61.}, they correspond to  elements of the
sheaf cohomology group $H^1(\Bbb{CP}'^{3|4},{\cal O})$. Such an
element is represented by a wavefunction which is a $(0,1)$-form
$\tilde\phi=d\bar Z^I\omega_{\bar I}(Z,\bar Z)$; it obeys
$\bar\partial\tilde\phi=0$, and is subject to the gauge
equivalence $\tilde\phi\to\tilde\phi+\bar\partial \alpha$, for any
function $\alpha$ on $\Bbb{CP}'^{3|4}$.  $\tilde\phi$ and the
gauge parameter $\alpha$ take values in the adjoint representation
of the gauge group, though we have not shown this in the notation.
Vertex operators for these states were described in \witten.

The relation between the two ways of describing gauge fields in
twistor space is described in section VI.5 of \ref\atiyah{M. F.
Atiyah, {\it Geometry Of Gauge Fields}, Lezioni Fermiane (Academia
Nazionale dei Lincei and Scuola Normale Superiore, Pisa, 1979).}
and is as follows. Suppose that $\phi$ is a function on
$\Bbb{RP}^{3|4}$. It is defined for real values of the ratio
$z=\lambda^2/\lambda^1$. We assume that $\phi$ is real-analytic
and so can be analytically continued to a neighborhood of
$\Bbb{RP}^{3|4}$ in $\Bbb{CP}^{3|4}$; moreover, we assume that
this neighborhood includes all points in $\Bbb{CP}'^{3|4}$ where
$z$ is real. Then we define $\tilde\phi=\phi\cdot
\bar\partial\left(\vartheta({\rm Im}\,z)\right)$ (where
$\vartheta({\rm Im}\,z)$ is equal to 1 for ${\rm Im}\,z>0$ and 0
for ${\rm Im}\,z<0$). The mapping $\phi\to\tilde\phi$ is the
mapping from vertex operators that describe gauge fields in the
open string approach to twistor-string theory to those that
describe vertex operators in the $B$-model of $\Bbb{CP}^{3|4}$.

Now let us consider the conformal supergravity sector.
 In the $B$-model, the most obvious closed string mode is a
deformation of the complex structure of $\Bbb{CP}'^{3|4}$. (The
$B$-model, after all, is used to describe complex structure
deformations in compactification of physical string theories on a
Calabi-Yau threefold.)  However, the deformation must preserve the
holomorphic volume-form or measure, which we will call $\Omega$,
since $\Omega$ is part of the definition of the $B$-model.  Let us
describe what sort of deformations have this property.  We cover
$\Bbb{CP}'^{3|4}$ with open sets $U_i$ which individually will be
undeformed.  (Explicitly, one can take two open sets, a set $U_1$
characterized by $\lambda^1\not= 0$ and  a set $U_2$ characterized
by $\lambda^2\not=0$.)  Then, one glues together the open sets
$U_i$ on their intersections $U_{ij}$ via diffeomorphism of the
form  $Z^I\to Z^I+\epsilon f_{ij}^I$, where (to describe the
deformation to first order) $\epsilon$ is an infinitesimal
parameter.  (Also, we define $f_{ji}=-f_{ij}$.) Here
$f_{ij}^I\partial/\partial Z^I$ is a holomorphic vector field --
so that the complex structures of $U_i$ and $U_j$ match together
on their intersection.  Moreover, so that the holomorphic measure
$\Omega$ can be defined on the deformed manifold, $f_{ij}$ must be
a {\it volume-preserving} vector field. This means explicitly that
\eqn\jonon{{\partial\over\partial Z^I}f_{ij}^I=0.} Finally, on
triple intersections $U_i\cap U_j\cap U_k$, compatibility of the
gluings requires that $f_{ij}+f_{jk}+f_{ki}=0$.  There are no such
triple intersections in our explicit covering of $\Bbb{CP}'^{3|4}$
by two open sets, so in that example this condition is trivial.
Taking all this together, the $f$'s describe an element of the
sheaf cohomology group $H^1(\Bbb {CP}'^{3|4},T')$, where $T'$ is
the sheaf of volume-preserving vector fields.

In $\bar\partial$ cohomology, an element of this cohomology group
is described by a wavefunction $\hat J=d\bar Z^{\bar I}j_{\bar
I}^K$ which obeys $\bar\partial \hat J=0$ (explicitly
$\partial_{\bar I}j_{\bar J}^K-\partial_{\bar J}j_{\bar I}^K=0$)
and is volume-preserving, that is $\partial_Kj_{\bar I}^K=0$.
$\hat J$ is subject to the usual gauge equivalence $\hat J\to \hat
J+\bar\partial\alpha$ for any section $\alpha$ of $T'$.  The
relation between $\hat J$ and the corresponding object $f^K$ in
the open string case is easy to guess, by analogy with what we
said in the gauge theory case: assuming $f^K$ has a sufficient
degree of analyticity, the relation is $d\bar Z^{\bar I}j_{\bar
I}^K=f^K\cdot\bar\partial( \vartheta({\rm Im}\,z))$.

Using the notation of eqn. (4.11) of \witten, a vertex operator
corresponding to $\hat J$ is \eqn\hutyy{V_{\hat J}=\eta^{\bar
I}j_{\bar I}^K\theta_K.} (The operator we have written is a
$(0,0)$-form; a $(1,1)$-form is obtained by the standard
``descent'' procedure.)  Here $\eta^{\bar I}$ and $\theta_K$ are
worldsheet fermions of the $B$-model. From this point of view, we
do not understand why $\hat J$ has to be volume-preserving.

The Penrose transform \ref\penroseb{R. Penrose, ``The Nonlinear
Graviton,'' Gen. Rel. Grav. {\bf 7} (1976) 171.} shows that a
volume-preserving deformation of the complex structure of twistor
space describes a solution of the anti-self-dual Weyl equations in
spacetime.  This describes one helicity of conformal supergravity.
Where does the other helicity come from?

It is plausible to postulate another type of closed string mode
with vertex operator $b_{\bar IK}$ that couples to $D1$-branes via
\eqn\furro{\int_{\Bbb{D}} b_{\bar I K}d\bar Z^{\bar I}\wedge
dZ^K,} where $\Bbb{D}$ is the world-volume of a $D1$-brane.  In
physical string theory, such a mode would arise in the RR sector
(and might be called an RR $B$-field).   On $b$, we can impose the
standard equation of motion and gauge invariance of the $B$-model,
$\bar\partial b=0$, $b\to b+\bar\partial \lambda$.  If these were
the only conditions, $b$ would be understood as an element of
$H^1(\Bbb{CP}'^{3|4},T^*)$, where $T^*$ denotes the cotangent
bundle.  The coupling \furro\ is invariant under the additional
transformation $b_{\bar IK}\to b_{\bar I K}+\partial_Kw_{\bar I}$.
We assume that $b$ is subject to this additional invariance. $b$
can then be related to the abelian gauge field $g$ of the open
string case by the familiar formula $d\bar Z^{\bar I}b_{\bar I
K}=g_K\cdot\bar\partial(\vartheta({\rm Im}\,z))$.

The vertex operator \hutyy\ is not automatically on-shell.  We
would like to postulate an effective action whose associated
Euler-Lagrange equation places $j$ on-shell. At the linearized
level, the appropriate action is
\eqn\yuru{\int_{\Bbb{CP}^{3|4}}d\bar X^{\bar I}d\bar X^{\bar
J}d\bar X^{\bar K}b_{\bar I I}\partial_{\bar J}j_{\bar K}^I
\Omega.} Here $X^I$ are local complex coordinates on
$\Bbb{CP}^{3|4}$ (as opposed to the homogeneous coordinates $Z^I$
used in most of our formulas), and  the complex conjugates $\bar
X^I$ are purely bosonic.  (Some of the $X^I$ are fermionic, but as
explained in \witten, there is no need to introduce complex
conjugates of the fermionic coordinates.) Upon varying with
respect to $b$,  \yuru\ leads to $\bar\partial j=0$ as an equation
of motion.  It is conceivable that one should introduce additional
fields so that the condition for $j$ to be volume-preserving would
also arise as an equation of motion; however, we do not know a
convenient way to do this.  We also do not know how to explicitly
show in the string theory the origin of the term \yuru\ in the
effective action.

While \yuru\ is adequate to linear order in $j$, we would like to
write a suitable action for complex structure deformations that is
not limited to linear order.  This can be done as follows.  The
``field'' in the action will be an almost complex structure, which
is a tensor $J^A{}_B$ constrained to obey $J^2=-1$.  The indices
$A$ and $B$ can be either holomorphic or antiholomorphic.  The
unperturbed $J$ is $J^I{}_J=i$, $J^{\bar I}{}_{\bar J}=-i$, with
other components vanishing. The first order perturbation of $J$,
subject to the constraints, has matrix elements $J^I{}_{\bar J}$
and $J^{\bar I}{}_J$; the components $J^I{}_{\bar J}$ are what we
have hitherto called $j_{\bar J}^I$.  From $J$ one can construct
an invariant tensor $N_{\bar I\,\bar J}{}^K$ called the Nijenhuis
tensor; the almost complex structure $J$ is called integrable if
and only if $N=0$. The linearized approximation to $N$ is $N_{\bar
I\,\bar J}{}^K=\partial_{\bar I}j_{\bar J}{}^K-\partial_{\bar
J}j_{\bar I}{}^K$.  The nonlinear extension of the action \yuru\
is \eqn\yturu{\int_{\Bbb{CP}^{3|4}}d\bar X^{\bar I}\,d\bar X^{\bar
J}d\bar X^{\bar K}b_{\bar I I}N_{\bar J\,\bar K}^I \Omega.} The
equation of motion for $b$ asserts that $N=0$, or in other words
the almost complex structure is integrable.

According to the Penrose transform \penroseb, the condition $N=0$
corresponds in Minkowski spacetime to $W_{abcd}=0$, where
$W_{abcd}$ and $W_{\ad \bd \cd \dot d}$, which are symmetric in
all their indices, are the self-dual and anti-self-dual parts of
the Weyl tensor. (To be more precise, in the case considered by
Penrose, the result is $W_{abcd}=0$, while in our situation, one
will get a supersymmetric extension of this.)
 This strongly suggests that the spacetime interpretation of
\yturu\ would be an action $\int d^4x \sqrt g \,U^{abcd} W_{abcd}$
(or rather a supersymmetric extension of this \ref\gates
%NBnew added reference
{S. Ketov, H. Nishino and S.J. Gates, Jr., ``Selfdual Supersymmetry
and Supergravity in Atiyah-Ward Space-Time,'' Nucl. Phys.
{\bf B393} (1993) 149, hep-th/9207042.}), where $U^{abcd}$
(symmetric in all its indices) is a field of Lorentz spin $(2,0)$,
just like $W^{abcd}$. As we discuss momentarily, $U$ is part of
the spacetime interpretation of the twistor field $b$.

To get conformal supergravity, we must, as in section 4 of
\witten, generate from $D$-instantons (which would correspond to
worldsheet instantons in the other approach to twistor-string
theory) a $U^2$ interaction. The resulting action \eqn\obvo{\int
d^4 x \sqrt g\left(U^{abcd} W_{abcd} -\half \epsilon U^2\right)}
is equivalent, after integrating out $U$, to ${1\over
2\epsilon}\int d^4x\sqrt g W^{abcd}W_{abcd}$, which is the action
of conformal gravity.

To be more exact, the standard conformal gravity action is
$${1\over 4\epsilon}\int d^4x\sqrt g\left( W^{abcd} W_{abcd}+
W^{\ad\bd\cd\dot d} W_{\ad\bd\cd\dot d} \right).$$ As in the gauge
theory case treated in \witten, the two differ by
$${1\over 4\epsilon}\int d^4x\sqrt{g}
\left( W^{abcd} W_{abcd}- W^{\ad\bd\cd\dot d} W_{\ad\bd\cd\dot d}
\right),$$ which is a topological invariant that does not affect
perturbation theory.

In the present situation, it is not hard to see where the $U^2$
term will come from.  We have already postulated the coupling
\furro\ of the field $b$ to $D1$-branes.  This coupling implies
that the contribution of a $D1$-brane is proportional to
$\exp(-\int_\Bbb{D} b)$. Expanding this in powers of $b$ and
integrating over moduli, the part of the effective action
quadratic in $b$ is \eqn\tunron{{1\over 2}\int_{\cal M}d\mu
\left(\int_\Bbb{D}b\right)^2,} from which we will extract the
$U^2$ term.  ${\cal M}$ denotes the component of the moduli space
of curves in $\Bbb{CP}^{3|4}$ that contains $\Bbb{D}$; the
corresponding contribution to the effective action is evaluated by
integrating over this moduli space with a suitable measure $d\mu$,
as noted in \tunron.

The relevant case is  the case that $\Bbb{D}$ is a degree one
instanton, that is a copy of $\Bbb{CP}^1$ linearly embedded in
$\Bbb{CP}^{3|4}$. In the Penrose transform, as reviewed in
\witten, such a $\Bbb{D}$ corresponds to a point in Minkowski
spacetime, or more exactly, a point in a chiral Minkowski
superspace with coordinates $x^{a\dot a}$, $\theta^{Aa}$. Thus, we
can define a function ${\cal W}(x,\theta)$ on Minkowski superspace
by ${\cal W}(x,\theta)=\int_{\Bbb{D}_{x,\theta}}b$, where
$\Bbb{D}_{x,\theta}$ is the curve with moduli $x$ and $\theta$.
(${\cal W}$ is holomorphic since $\bar\partial b=0$.) The $\theta$
expansion of ${\cal W}$ is discussed in more detail in section
4,\foot{In section 4, the Weyl tensor appears instead of $U$, as
the expansion is made in a framework in which $U$ has been
integrated out.} but for now we simply note that it reads in part
\eqn\hyto{{\cal W}=C+\dots +{1\over
4!}\epsilon_{ABCD}\theta^{Aa}\theta^{Bb}\theta^{Cc}\theta^{Dd}U_{abcd}+\dots}
Here $U$ is the field we want, and $C$ is a scalar that will turn
out to be a chiral ``dilaton.''

The interaction \tunron\ becomes \eqn\ononps{\int d^4x^{a\dot
a}d^8\theta^{Aa}\, {\cal W}^2.} Upon performing the theta integral
using \hyto, we do get the desired $\int d^4x \sqrt g U^2$
coupling.

We can also now see that $C$ behaves as a dilaton.  Let $k$ be a
Kahler form of $\Bbb{CP}^{3|4}$, normalized so that for a degree
one curve $\Bbb{D}$, $\int_\Bbb{D}k=1$.  Consider shifting $b$ by
$b\to b+ck$, where $c$ is a complex constant.  This shifts the
scalar field $C$ by a constant, $C\to C+c$.  Now, let $\Bbb{D}'$
be any $D$-instanton of degree $d$.  Then (by the definition of
the degree) $\int_{\Bbb{D}'}k=d$, so under $C\to C+c$,
$\int_{\Bbb{D}'}k\to \int_{\Bbb{D}'}k+dc$.  It follows that under
this shift, the contribution of $\Bbb{D}'$ to a scattering
amplitude, which is proportional to $\exp(-\int_{\Bbb{D}'}b)$, is
multiplied by $\exp(-dc)$. Differently put, if $\langle C\rangle$
denotes the expectation value of $C$, then the $\langle C\rangle$
dependence of a degree $d$ contribution to the scattering
amplitudes is \eqn\toffogo{\exp(-d\langle C\rangle).}

This is not the complete story. Parity invariance implies that
there must be another scalar field $\bar C$ with parity conjugate
couplings; for example, $\bar C$ couples to $W_{\ad\bd\cd\dot
d}^2$ while $C$ couples to $W_{abcd}^2$.  $\bar C$ is the ``top''
component of the superfield ${\cal W}$, as we describe more fully
in section 4.

The results that we have just described have counterparts in the
open string approach to twistor-string theory. In that context,
the form of the vertex operator $V_g=g_I \partial Z^I$ shows that
there is a coupling $\int_{\partial {D}}g_IdZ^I$ of the $g$-field
to the boundary $\partial{D}$ of an open string worldsheet ${D}$.
If this boundary is a ``line'' $\Bbb{D}$ in $\Bbb{RP}^{3|4}$,
corresponding to a point in real Minkowski superspace (of $++--$
signature), then the definition of the corresponding superspace
field is ${\cal W}=\int_{\Bbb{D}}g_IdZ^I$.

\newsec{Minkowski Space Interpretation}

In this section, we determine the spectrum of massless fields in
Minkowski spacetime that is associated with the twistor space
vertex operators found in section 2.  It does not matter which
type of twistor-string theory we use, since the two types of
wavefunction are related by a map ($\phi\to
\phi\cdot\bar\partial(\vartheta({\rm Im}\,z))$) that was described
in section 2.  For brevity, we will use the open string language
in this section.

The basic input we need is that \refs{\penrose,\atiyah} a function
of the homogeneous coordinates $Z^I$ of twistor space that is
homogeneous in the $Z^I$ of degree $k$ describes a massless state
in Minkowski spacetime of helicity $1+k/2$.

As an example, let us consider the field $f^I(Z)$ found in section
2.  This field is a function of bosonic and fermionic variables
$\lambda^a,\mu^{\dot a}$, and $\psi^A$; let us first determine the
fields we get if we set $\psi^A=0$.

For each value of $I$, $f^I(\lambda,\mu)$ is homogeneous in $Z$ of
degree 1. So if we ignore the spin carried by the $I$ index, we
get four bosonic and four fermionic helicity states, each of
helicity 3/2.

Of course, it is not correct to ignore the spin carried by the $I$
index.  The possible choices of $I$ are $(\alpha,\dot\alpha, A)$,
where the cases $I=\alpha$ or $\dot\alpha$ are bosonic states, and
$I=A$ are fermionic states.  Both $\alpha$ and $\dot\alpha$ take
two possible values and, under rotation around any given direction
in space (the relevant direction is the direction of motion of a
massless state in Minkowski spacetime corresponding to $f^I(Z)$),
these two states have helicity $1/2$ and $-1/2$.  So from
$f^I(\lambda,\mu)$, before taking account of the constraint and
gauge-invariance, we get two bosonic states of helicity 2 and two
of helicity 1.  As for the fermions, since the index $A$ carries
no helicity but transforms as ${\bf 4}$ of the $SU(4)$ group of
$R$-symmetries, we get four states of helicity $3/2$ and
transforming in that representation.

The gauge invariance $f^I\to f^I+Z^I\Lambda$ tells us to discard
the state described by a function $\Lambda(\lambda,\mu)$ that is
homogeneous of degree zero -- in other words, a bosonic state of
helicity 1.  The constraint $\partial_If^I=0$ likewise removes the
state described by the function $\partial_If^I$, which is
homogeneous of degree zero and so describes another bosonic state
of helicity 1.  After removing these two, we are left with two
bosonic states of helicity 2 and four helicity $3/2$ fermions
transforming in the ${\bf 4}$ of $SU(4)_R$.

Here is another way to do this counting for the bosonic states.
The two functions $f^a$ have the same content as the  two
Lorentz-invariant functions $\lambda_af^a$ and $\partial
f^a/\partial\lambda^a$, and likewise the $f^{\dot a}$ are
equivalent to two more Lorentz-invariant functions $\mu_{\dot
a}f^{\dot a}$ and $\partial f^{\dot a}/\partial \mu^{\dot a}$.
These two functions of degree 2 and two of degree 0 again describe
two bosonic states of helicity 2 and two of helicity 1. The two of
helicity 1 are removed, as before, using the gauge invariance and
constraint.

This gives us the states at $\psi=0$.  To get the full spectrum,
we expand in powers of $\psi$:
$f^I(\lambda,\mu,\psi)=f^I_0(\lambda,\mu)+f^I_{1\,A}(\lambda,\mu)\psi^A
+f^I_{2\,AB}(\lambda,\mu)\psi^A\psi^B+\dots.$ Here $f^I_k$ is
homogeneous in $\lambda,\mu$ with degree $1-k$, and so describes a
massless state of helicity $3/2-k/2$ if we ignore the angular
momentum carried by the $I$ index.  Upon taking that angular
momentum into account, as well as the gauge-invariance and the
constraints, we get the full collection of helicity states
described by the field $f^I(Z)$: \eqn\fsuper{ \eqalign{\lambda^a
f_a:~ & (2,{\bf 1}), ({3/ 2},{\bf{{\bar 4}}}), (1, {\bf 6}), (1/2,
{{\bf 4}}), (0,{\bf 1})\cr \mu^{\dot a}f_{\dot a}:~&(2,{\bf 1}),
({3/ 2},{\bf{{\bar 4}}}), (1, {\bf 6}), (1/2, {\bf 4}), (0,{\bf
1})\cr
 f^A:~& (3/ 2, {\bf 4}), (1,{\bf 15}\oplus {\bf 1}), (1/2,
{\bf{\bar{20}}}\oplus {\bf{\bar 4}}), (0, {\bf 10}\oplus {\bf 6})
, (-1/2, {\bf 4}).\cr}} Here the first entry is the helicity, and
the second  is the $SU(4)_R$ representation.

Next, we perform a similar analysis for $g_I(Z)$, first
considering the fields that arise at $\psi^A=0$. The Lorentz
scalars \eqn\scalarstwo{(Z^a g_a,~ Z^\ad g_\ad,~ \p_a g^a,~ \p_\ad
g^\ad)} are homogeneous of degrees \eqn\charge{(0,0,-2,-2).} To
allow for the gauge-invariance $g_I\to g_I+\partial_I\Lambda$ and
the constraint $Z^Ig_I=0$, we should remove the two fields of
degree 0. So we are left with two twistor space fields of degree
$-2$, describing states in Minkowski space of helicity 0. And the
fields $g_A$ are homogeneous of weight $-1$ and therefore describe
massless fields with helicity $+1/2$.

Allowing for the dependence on $\psi^A$, the complete massless
superfields described by $g_I$ are \eqn\gsuper{\eqalign{ \p_a g^a
:~& (0,{\bf 1}), (-1/2,{\bf \bar 4}), (-1, {\bf 6}), (-{3/ 2},
{\bf{4}}), (-2,{\bf 1})\cr \p_\ad g^\ad:~& (0,{\bf 1}), (-{1/
2},{\bf{\bar 4}}), (-1,{\bf{ 6}}), (-{3/ 2}, {\bf{4}}),
(-2,{\bf{1}})\cr g_A:~& ({1/ 2}, {\bf \bar 4}),
(0,{\bf{\bar{10}\oplus  6}}), (-1/2,{\bf{20\oplus 4}}),
(-1,{\bf{15\oplus 1}}),(-{3/ 2}, {\bf{\bar 4}}).\cr}}

The above results show that the massless fields described by $g_I$
have the opposite helicities and conjugate SU(4) representations
from those described by $f^I$, permitting these fields to be
combined together in writing an action, as proposed in section
2.2.  This duality between $f$ and $g$ can be understood by
defining the Fourier-like transform\foot{The argument for parity
symmetry in \ref\berkmotl{N. Berkovits and L. Motl, ``Cubic
Twistorial String Field Theory,'' JHEP {\bf 0404} (2004) 056,
hep-th/0403187.} involved a stringy extension of this Fourier
transform.}
 \eqn\fourier{ \widetilde g^I(V) =
\int_{\Bbb{RP}^{3|4}} d\Omega \exp({Z^K V_K}) f^I(Z).} Here the
$V_K$ are homogeneous coordinates on a new $\Bbb{RP}^{3|4}$ that
is dual to the original one.   The integral over $\Bbb{RP}^{3|4}$
is defined using the scaling-invariant measure $d\Omega$ (roughly
$ \l^a d\l_a \,d^2 \mu \, d^4\psi$) described in \witten.\foot{Our
notation is slightly non-standard; we write $\Omega$ for the
``volume-form'' on twistor space and, when convenient, we write
$d\Omega$ for the associated measure on real twistor space.} Given
that $f^I$ is homogeneous of degree 1 in $Z^I$, $\widetilde g^I$ is
homogeneous of degree $-1$ in $V_I$. Moreover, the Fourier
transform maps the gauge-invariance $\delta f^I=Z^I\Lambda$ to a
gauge-invariance $\delta \widetilde g^I=\partial^I\Lambda$, and the
constraint $\partial_If^I=0$ to a constraint $V_I\widetilde g^I=0$. In
other words, the Fourier transform maps the $f^I$ field in the
original $\Bbb{RP}^{3|4}$ to a dual field $\widetilde g^I$ on the dual
$\Bbb{RP}^{3|4}$.  Since the Fourier transform is a
$PSU(4|4)$-invariant operation, it is clear that $f^I$ describes
spacetime fields with the same quantum numbers as those described
by $\widetilde g^I$.  Because the $V_I$ have quantum numbers dual to
those of the $Z^I$, $\widetilde g^I$ describes states with quantum
numbers dual to those described by $g_I$, so $g_I$ and $f^I$
describe states with dual quantum numbers, as we have seen more
laboriously above.

\bigskip\noindent{\it Lack Of Unitarity}

So far we have exploited Lorentz-invariance in constructing the
scalar  functions $\sigma=\lambda_a f^a$ and $\sigma'=\mu_{\dot
a}f^{\dot a}$.   But we have not considered the rest of the
Poincar\'e group.  How, in fact, do the states transform under
spacetime translations?

A twistor space function of definite homogeneity describes a
particle state in Minkowski spacetime of definite helicity.
Spacetime translations act on the Minkowski coordinates $x^{a\dot
a}$ in the familiar fashion $x^{a\dot a}\to x^{a\dot a}+c^{a\dot
a}$.  In terms of the $Z^I$, the transformation is
\eqn\juki{\mu^{\dot a}\to \mu^{\dot a}+c^{a\dot a}\lambda_a.} The
generator of this transformation is \eqn\uncuc{D=c^{a\dot
a}\lambda_a{\partial\over\partial\mu^{\dot a}}.}

To determine how a vector field $f^I\partial_I$ commutes with
translations, we must evaluate the commutator $[D,f^I\partial_I]$.
When we do this, we get two kinds of term.  One term arises from
$[D,f^I]$.  This term describes the action of translations on
$f^I$, as if the components of $f^I$ were scalar functions on
twistor space of the appropriate homogeneity.  Let us write $P$
for the operator that maps $f^I\partial_I$ to $[D,f^I]\partial_I$.
There is also a second term which arises because $\partial_I$ does
not commute with $D$.  This second term arises precisely if $I=a$,
since $\partial/\partial\lambda^a$ does not commute with $D$ (but
 $\partial/\partial\mu^{\dot a}$ and $\partial/\partial\psi^A$ do so
commute).

The net effect is that on the pair $\left(\matrix{f^{\dot a}\cr
f^{ a}\cr}\right)$, the translation generator $D$ acts as
\eqn\nonnnop{\left(\matrix{P & * \cr 0 & P \cr}\right),} where $P$
would represent ordinary translations and the off-diagonal term
$*$ arises from $[D,\partial_I]\not=0$.

The matrix \nonnnop\ is not diagonalizable.  This clashes with our
usual experience.  We are accustomed to the idea that the
translation generators are Hermitian operators and so can be
diagonalized.  However, conformal supergravity is not a unitary
theory, and one symptom of this is that the translation generators
are undiagonalizable.

Clearly, on vector fields $f^{\dot a}\partial_{\dot a}$,
corresponding to vertex operators $f^{\dot a}Y_{\dot a}$, the
translation generators can be diagonalized. The same is true for
vertex operators $f^AY_A$.  They correspond to plane waves in
Minkowski spacetime. The spacetime interpretation of the other
vertex operators $f^aY_a$ will become clearer in the next section.

A similar analysis for the dual fields $g_I\partial Z^I$ shows
that on the pair $\left(\matrix{g_{\dot a}\cr g_{ a}\cr}\right)$,
the translations act as the transpose of \nonnnop.  Hence the
translations can be diagonalized in acting on vertex operators
$g_a\partial Z^a$.

\newsec{Spacetime Interpretation}

In this section, we first review the linearized description of
conformal supergravity in superspace, \nref\siegone{W. Siegel,
``On-shell O(N) Supergravity in Superspace,'' Nucl. Phys. {\bf
B177} (1981) 325.}%
\nref\berg{E. Bergshoeff, M. de Roo and B. de Wit, ``Extended
conformal supergravity,'' Nucl. Phys. {\bf B182} (1981) 173.}%
then determine the corresponding spectrum of massless helicity
states, and finally compare to the results of sections 2 and 3
based on twistor-string theory.

\subsec{Linearized Conformal Supergravity In Superspace}

At the linearized level, ${\cal N}=4$ conformal supergravity can
be described off-shell \refs{\siegone,\berg} by a  chiral scalar
superfield ${\cal W}(x^{a\dot a},\theta^A_a,\bar\theta^{\dot
a}_A)$ which satisfies the condition \eqn\reality{ \e^{ABCD} D^a_C
D_{Ea} D_{b D} D^b_F {\cal W} = \e_{EFGH} \bar D^{\ad A} \bar
D^G_\ad \bar D_\bd^B \bar D^{\bd H} \bar {\cal W}. } Here $D^a_A$
and $\bar D_{\dot a}^A$ are the usual superspace derivatives; it
is convenient to choose coordinates with $D^a_A =
{\p\over{\p\t_a^A}} +\bar\t_A^\ad \p_{a\ad}$, $\bar D^A_\ad =
{\p\over{\p\bar\t_A^\ad}} $, so that the condition for ${\cal W}$
to be chiral, namely $\bar D^A_\ad {\cal W}=0$, reduces to the
statement that ${\cal W}$ is independent of $\bar\t^\ad_A$. The
field ${\cal W}$ has an analog in ${\cal N}=2$ super Yang-Mills
theory; that theory is described \ref\whot{R. Grimm, M. Sohnius
and J. Wess, ``Extended Supersymmetry and Gauge Theories,'' Nucl.
Phys. {\bf B133} (1978) 275.}
%NB added reference
at the linearized level by an adjoint-valued chiral superfield
${\cal W}_{YM}$ satisfying \eqn\realym{ D^a_C D_{Da} {\cal W}_{YM}
= \bar D^{\ad }_C \bar D_{D\ad} \bar {\cal W}_{YM} .} These
conditions are customarily called reality conditions because they
do in fact imply that certain components in the $\theta$
expansions of ${\cal W}$ and ${\cal W}_{YM}$ are real, while
others obey Bianchi identities. For example, \realym\ imples that
the auxiliary fields $d_{(CD)} =D^a_C D_{D a}{\cal W}_{YM}$ are
real and that the  Yang-Mills field strength, whose self-dual part
is $F_{ab}=\e^{AB} D_{A a} D_{B b} {\cal W}_{YM}$, satisfies the
usual Bianchi identities. Similarly, \reality\ implies that the
auxiliary fields $d_{[EF]}^{[AB]} = \e^{ABCD} D^a_C D_{Ea} D_{b D}
D^b_F {\cal W} $ are real and that the Weyl  tensor, whose
self-dual part is  $W_{abcd}=\e^{ABCD} D_{Aa} D_{Bb} D_{Cc} D_{Dd}
{\cal W} $, satisfies the usual Bianchi identities.

The component field expansion of ${\cal W}$ is
\eqn\compw{\eqalign{{\cal W}(x,\t)=& C + \t^{Aa} \Lambda_{A a} +
(\t^2)^{(AB)} E_{(AB)} + (\t^2)^{(ab)}_{[AB]} T^{[AB]}_{(ab)}
+(\t^3)^{(abc)}_D (\p\eta)^D_{(abc)} \cr
&+ (\t^3)^{aC}_{[AB]}
\xi^{[AB]}_{aC} + (\t^4)^{A(ab)}_B (\p V)_{(ab)A}^B +
(\t^4)^{(abcd)} W_{abcd} +(\t^4)_{[CD]}^{[AB]}
d^{[CD]}_{[AB]} \cr & +
 (\t^5)^{a[AB]}_C \p_{a \ad}\bar\xi^{\ad C}_{[AB]}
+ (\t^5)^{A(abc)} (\p\rho)_{A(abc)} +(\t^6)_{(AB)}  \p_\mu\p^\mu
\bar E^{(AB)} \cr & +(\t^6)_{[AB](ab)} \p^{\a\ad}\p^{\b\bd}\bar
T_{(\ad\bd)}^{[AB]} +(\t^7)_{A a} (\p^\mu \p_\mu)
\p^{\a\ad}\bar\Lambda_\ad^A +(\t^8) (\p_\mu\p^\mu)^2 \bar C.\cr}}
By virtue of \reality, the component fields in this expansion obey
various conditions.  $W_{abcd}$, which in the nonlinear theory is
interpreted as the self-dual Weyl tensor, obeys Bianchi identities
which imply that it can constructed from a vielbein
$e_\mu^{a\ad}$.  $d$ obeys $d^{[BC]}_{[AB]}=0$, as well as being
real, and similarly $\xi_{[AB]}^B=0$.  Bianchi identities
following from \reality\ have been used to write certain
components of the $\theta$ expansion of ${\cal W}$ in terms of
potentials $\eta$, $V$, $\rho$; here, $V^A_{\mu A}=0$ and
\eqn\trunon{\eqalign{ (\p V)_{(ab)A}^B =& \p_\mu V_{\nu A}^B
(\s^{\mu\nu})_{(ab)},\cr (\p \eta)_{(abc)}^A =& \p_\mu \eta_{\nu
(a}^A (\s^{\mu\nu})_{bc)},\cr(\p \rho)_{(abc)A} =& \p_\mu
\rho_{\nu A (a} (\s^{\mu\nu})_{bc)}.\cr}}  Moreover, $\rho_{\mu A
a}$ is related to $\bar\eta_{\mu A \ad}$ by the formula
\eqn\rhodef{\rho_{\nu A a}= \s^\mu_{\a\ad}(\p_\mu\bar\eta_{\nu
A}^\ad -\p_\nu\bar\eta_{\mu A}^\ad +\half
\e_{\mu\nu\tau\kappa}\p^\tau \bar\eta^{\kappa \ad}_A).}

The linearized action for ${\cal N}=4$ conformal supergravity is
\eqn\linact{S = \int d^4 x \int d^8 \t \,\,{\cal W}^2.} Upon
performing the $\theta$ integrals, this gives the component action
\eqn\compact{\eqalign{S = \int d^4 x &\left( C (\p^\mu \p_\mu)^2
\bar C + \Lambda_{a A} (\p^\mu \p_\mu) \p^{\a\ad}
\bar\Lambda^A_\ad + E_{(AB)} \p_\mu\p^\mu \bar E^{(AB)} \right.\cr
&\left.+ T^{[AB]}_{(ab)} \p^{\a\ad}\p^{\b\bd} \bar
T_{(\ad\bd)[AB]} + \xi_{a C}^{[AB]} \p^{\a\ad} \bar\xi_{\ad[AB]}^C
+ d^{[AB]}_{[CD]} d^{[CD]}_{[AB]}\right.\cr & \left. + (\p
V)_{(ab) A}^B (\p V)^{(ab) A}_B+ R_{(abcd)} R^{(abcd)} +
(\p\eta)^A_{(abc)} (\p\rho)_A^{(abc)}\right).}}

\subsec{Spectrum of Conformal Supergravity}

To find the helicities described by these fields, one needs to
analyze solutions to their higher-derivative equations of motion.
Though this is elementary and not essentially novel (see
\ref\zumino{S. Ferrara and B. Zumino, ``Structure of Linearized
Supergravity and Conformal Supergravity,'' Nucl. Phys. {\bf B134} (1978)
301.}), the details are slightly unfamiliar, because of the higher
derivatives appearing in the kinetic operators.

For example,    the equation of motion for the scalar field $C(x)$
is $\square^2C=0$, where $\square=\partial_\mu\partial^\mu$ is the
usual kinetic operator for a massless scalar field.  The equation
$\square C=0$ has for its general solution a superposition of
plane waves $\exp(ik\cdot x)$, where $k^2=0$. Being of fourth
order instead of second order, the equation $\square^2C=0$ must
have a general solution that depends on twice as many functions.
Indeed, this equation is obeyed by the plane wave
$\sigma_k=\exp(ik\cdot x)$, but also by $\sigma_k'=A\cdot
x\exp(ik\cdot x)$, for any vector $A$.  It may seem that we now
have too many solutions, because of the choice of $A$.  However,
if $A\cdot k=0$, then $\sigma'_k$ is a linear combination of the
derivatives of $\sigma_k$ with respect to $k$, and for purposes of
constructing the general solution of $\square^2C=0$, since we will
include arbitrary linear combinations of the $\sigma_k$ anyway,
such choices of $A$ are irrelevant. So for enumerating the
possible solutions, we pick an $A$ with $A\cdot k\not= 0$, and the
precise choice of $A$ does not matter.

If now we consider how translations act on the pair of functions
$\sigma_k$, $\sigma'_k$, we obtain a result that is in precise
parallel with what we described from the twistor point of view at
the end of section 3.  If the translation operator $P_{a\dot
a}=-i\partial/\partial x^{a\dot a}$ would act only on the plane
waves $\exp(ik\cdot x)$ in the definition of $\sigma_k$ and
$\sigma'_k$, it would multiply those functions by $k_{a\dot a}$;
because of the prefactor $A\cdot x$ in the definition of
$\sigma'_k$, $P_{a\dot a}$ actually acts on the pair
$\left(\matrix{\sigma_k\cr \sigma'_k\cr}\right)$ via the matrix
\eqn\onono{\left(\matrix{k_{a\dot a} &-iA_{a\dot a}\cr 0 &
k_{a\dot a}\cr}\right).} This matrix is undiagonalizable, which as
we noted in section 3 reflects the nonunitarity of the theory.

Since the choice of $A$ is arbitrary, we can take it to be a unit
vector in the time direction.   Making this choice, we write the
general solution of the equation $\square^2C=0$ as \eqn\gensol{
\eqalign{C(x)& = \int d^3 k \,e^{ik\cdot x + i|k|t}
\left(C_0(|k|,k) + {it\over |k|} C'_{0}(|k|,k))\right) \cr & +\int
d^3 k \,e^{ik\cdot x - i|k|t} \left(C_0(-|k|,k) - {it\over |k|}
C'_{0}(-|k|,k)\right).\cr}} As in the case of an ordinary massless
scalar field, this can be written more conveniently as
\eqn\csol{C(x) = \int d^4 k \d(k^2) e^{ik\cdot x} \left(C_0(k) +
i{x_0\over k_0} C'_{0}(k)\right),} where $C_0(k)$ and $C'_{0}(k)$,
defined only for $k^2=0$, are independent  fields of helicity
zero.  They make up what is called a ``dipole.''  Note that the
normalization of $C'_{0}$ has been chosen such that $\square C =
\int d^4 k \d(k^2) e^{ik\cdot x} C'_{0}(k).$

The linearized action of conformal supergravity also contains a
spinor field $\Lambda_A^a$ with the third order equation of motion
$\square\p_{a\ad} \Lambda^a_A=0.$  We would like in a similar
fashion to describe the helicity states of such a spinor field; we
do this by describing the solutions of the equation whose
wavefunctions are $\exp(ik\cdot x)$ times a polynomial.  These are
the states that have momentum $k$ in the sense that they form a
space (analogous to the space spanned by $\sigma_k$ and
$\sigma'_k$) in which the only eigenvalue of the translation
operator is $k$.

To analyze these ``plane waves,'' we introduce a pair of spinors
$\pi,\tilde\pi$ such that $k^{a\dot a}=\pi^a\tilde\pi^{\dot a}$,
and a second pair of  spinors $\tau,\tilde\tau$ such that
\eqn\membo{\pi^a\tilde\tau_a =1,\quad \tilde\pi^\ad \tau_\ad =1.}
In particular, $\pi$ and $\tilde \tau$ give a basis for the space
of positive chirality spinors, so we can expand
$\Lambda^a_A = \pi^a \Lambda^{(1)}_A + \tilde\tau^a
\Lambda^{(2)}_A.$ Since $\square\p_{a\ad} \Lambda^a_A=0$ implies
%\eqn\jukop{\Lambda^a_A = \pi^a \Lambda^{(1)}_A + \tilde\tau^a
%\Lambda^{(2)}_A.} Since $\square\p_{a\ad} \Lambda^a_A=0$ implies
that $\square^2 \Lambda^{(1)}_A=0$ and that $\square
\Lambda^{(2)}_A=0$, the most general solution is
\eqn\lambdasol{\Lambda_A^{a}(x) = \int d^4 k \d(k^2) e^{ik\cdot x}
\left(\pi^a (\Lambda_{-\half A}(k) + i{{x_0}\over
{k_0}}\Lambda'_{-\half  A}(k)) +\tilde\tau^a \Lambda_{\half
A}(k)\right).} The plane wave $\pi^a\e^{ik\cdot x}$ has helicity
$-1/2$ (it is the standard plane wave solution of the ordinary
chiral Dirac equation $\partial_{a\dot a}\Lambda^a=0$!), and
$x_0/k_0$ is invariant under spatial rotations.  So the terms
$\Lambda_{-\half A}$ and $\Lambda'_{-\half A}$ describe waves of
helicity $-1/2$.  As $\tilde\tau$ transforms under rotations
around the $\vec k$ axis oppositely to $\pi$, $\Lambda_{\half A}$
describes a wave of helicity $+1/2$.
 $\Lambda_{-\half A}$ and $\Lambda'_{-\half  A}$ combine to a
 ``dipole''
field of helicity $-\half$, on which spacetime translations act in
a nondiagonalizable fashion as in \onono. On the other hand,
$\Lambda_{\half A}$ transforms as an ordinary field of helicity
$+\half$.

For $T_{(ab)}^{[AB]}$, the linearized equation of motion is
$\p^{\a\ad}\p^{\b\bd} T_{(ab)}^{[AB]}=0$.  This equation similarly
implies that \eqn\Tsol{T^{[AB]}_{(ab)}= \int d^4 k \d(k^2)
e^{ik\cdot x} \left(\pi_a\pi_b (T^{[AB]}_{-1}(k) + i{{x_0}\over
{k_0}}T'^{[AB]}_{-1 }(k)) +\pi_{(a} \tilde\tau_{b)}
T^{[AB]}_0(k),\right)} where on the right the subscript on
$T^{[AB]}$ denotes the helicity.

The linearized equations obeyed by the gravitino $\eta_{\mu a}^A$
and the graviton $e_{\mu a \ad}$ are \eqn\graveq{\p^{e\bd}
\p^{a(\dot c} \p^{\ad )b} \s^\mu_{b\bd}\eta_{\mu a}^A=0, \quad
\p^{\ad (c} \p^{d)\dot b} \p^{a(\dot c}\p^{\dot d) b}
\s^\mu_{b\bd}e_{\mu a\ad}=0.} By an analysis similar to the above,
these equations imply that \eqn\etasol{\eqalign{\eta^{A}_{\mu
a}=&\s_\mu^{b\bd} \int d^4 k \d(k^2) e^{ik\cdot x}
\left(\pi_a\pi_b\tau_\bd (\eta^{A}_{-{3\over 2}}(k) + i{{x_0}\over
{k_0}} \eta^{A}_{-{3\over 2} '}(k) ) \right.\cr
&+\left.\pi_{(a}
\tilde\tau_{b)}\tau_\bd \eta^{A}_{-\half} (k) +\tilde\tau_a
\tilde\tau_b\tilde\pi_\bd \eta^{A}_{{3\over 2}} (k)\right),\cr}} and
that \eqn\esol{\eqalign{e_{\mu a \ad}=&\s_\mu^{b\bd} \int d^4 k
\d(k^2) e^{ik\cdot x} \left(\pi_a\pi_b\tau_\ad\tau_\bd (e_{-2}(k)
+ i{{x_0}\over {k_0}} e_{-2 '}(k) ) +\pi_{(a}
\tilde\tau_{b)}\tau_\ad\tau_\bd e_{-1} (k)\right.\cr
&+\left.\tilde\tau_a \tilde\tau_b\tilde\pi_{(\ad}\tilde\tau_{\bd)}
e_1(k) + \tilde\tau_a\tilde\tau_b\tilde\pi_\ad\tilde\pi_\bd
(e_{2}(k) + i{{x_0}\over {k_0}} e_{2 '}(k))\right).\cr}} To obtain
this result, we have used the fact that polarizations of
$\eta_{\mu a}^A$ and $e_{\mu a\ad}$ which involve both $\pi_a$ and
$\tilde\pi_\ad$ can be gauged away using the gauge transformations
$$\d\eta_{\mu a}^A = \s_\mu^{\b\bd}(\pi_b\tilde\pi_\bd \Omega_a^A + \e_{ab}
\k_\bd^A), \quad \d e_{\mu a\ad}= \s_\mu^{b\bd}(\pi_b\tilde\pi_\bd
 \Sigma_{a\ad} + \e_{ab}\tilde l_{\ad\bd} +
\e_{\ad\bd} l_{ab}),$$ where $(\Omega_a^A, \k_\bd^A,\Sigma_{a\ad},
\tilde l_{\ad\bd}, l_{ab})$ are gauge parameters.

Finally, the equations of motion for $E_{(AB)}$, $\xi_{a C}^{[AB]}$,
and $V_{\mu A}^B$ are the usual equations for fields of helicity $0,\half$
and $1$.

$$\vbox{\hsize=4.5in\noindent
Table 1. $U(1)$ charges, helicities, and $SU(4)_R$ representations
of physical states in ${\cal N}=4$ conformal supergravity in four
dimensions.   Whenever a field gives rise to two states with the
same helicity, they form a ``doublet,'' with the translation
generators not being diagonalizable.
\smallskip
\halign to 4.5in{#\hfil\hfil\tabskip=5em plus5em
minus5em&\hfil\hfil#\tabskip=5em plus5em
minus5em&\hfil#\hfil\tabskip=5em plus5em
minus5em&\hfil#\hfil\tabskip=0pt\cr
%\multispan2\hfil Table Title\hfil\cr
\noalign{\medskip\hrule\medskip}  %\smallskip\hrule\medskip}
&$U(1)$ Charge  & Helicity & $SU(4)$ Representation\cr
\noalign{\medskip\hrule\medskip} $C$&$4$&$0,0$&$\bf 1$\cr
\noalign{\smallskip\smallskip}
$\Lambda_A^a$&$3$&$-{1\over2},-{1\over2}, {1\over2}$&$\bf
\bar{4}$\cr \noalign{\smallskip\smallskip} $E_{(AB)}$&$2$&$0$&$\bf
\bar{10}$\cr \noalign{\smallskip\smallskip}
$T_{(ab)}^{[AB]}$&$2$&$-1, -1, 0$&$\bf 6$\cr
\noalign{\smallskip\smallskip}
$\xi_{aC}^{[AB]}$&$1$&$-{1\over2}$&$\bf 20$ \cr
\noalign{\smallskip\smallskip} $\eta_\mu^{A a}$&$1$&$-{3\over2},
-{3\over2}, -{1\over2}, {3\over2}$&$\bf 4$ \cr
\noalign{\smallskip\smallskip} $V_{\mu A}^B$&$0$&$1,-1$&$\bf 15$
\cr \noalign{\smallskip\smallskip} $d_{[AB]}^{[CD]}$&$0$&{\rm
none}&$\bf 20$ \cr \noalign{\smallskip\smallskip} $e_{\mu}^{a \dot
a}$&$0$&$2, 2, 1, -1, -2, -2$&$\bf 1$\cr
\noalign{\smallskip\smallskip} $\bar{\eta}_{\mu A}^{\dot
a}$&$-1$&${3\over2}, {3\over2}, {1\over2}, -{3\over2}$&$\bf
\bar{4}$\cr \noalign{\smallskip\smallskip} $\bar{\xi}^{{\dot
a}C}_{[AB]}$&$-1$&${1\over2}$&$\bf \bar{20}$ \cr
\noalign{\smallskip\smallskip} $\bar{T}_{({\dot a}{\dot
b})}^{[AB]}$&$-2$&$1, 1, 0$&$\bf 6$\cr
\noalign{\smallskip\smallskip} $\bar{E}^{(AB)}$&$-2$&$0$&$\bf 10$
\cr \noalign{\smallskip\smallskip} $\bar{\Lambda}^{A}_{{\dot
a}}$&$-3$&${1\over2}, {1\over2}, -{1\over2}$&$\bf 4$\cr
\noalign{\smallskip\smallskip} $\bar{C}$&$-4$&$0,0$& $\bf 1$ \cr
\noalign{\medskip\hrule} }}
$$

Combining these results, we summarize in the table the helicities
and $SU(4)_R$ representations of the physical states of ${\cal
N}=4$ conformal supergravity.  In presenting the table, we have
also included a $U(1)$ $R$-charge, defined so that ${\cal W}$ has
charge $4$ and $\theta$ has charge $1$.  The $U(1)$ charges can be
read off from the $\theta$ expansion of eqn. \compw.  The analog
of this $U(1)$ charge in Yang-Mills theory is called $S$ in
\witten.  The $U(1)$ charge is not conserved by the interactions
of conformal supergravity. In twistor-string theory, it is
somewhat natural to add a constant to the $U(1)$ generator, so
that ${\cal W}$ has charge 0 while $\theta$ still has charge $1$.
Using this convention, which is adopted in \witten, the linearized
action (before integrating out auxiliary fields) has $S$ charge
$-4$, and $D$-instantons of degree $d$ and genus $g$ carry $S$
charge $-4(1+d-g)$. However, in the table we have simply defined
the $U(1)$ to be a symmetry of the linearized action.

 \subsec{ Identification With Twistor Fields}

We can now verify that the gauge-singlet sector of twistor-string
theory has the same physical states as conformal supergravity.
Indeed, the results of the table for the physical states of ${\cal
N}=4$ conformal supergravity in four dimensions agree with the
results from eqns. \fsuper\ and \gsuper\ for the spacetime fields
described by the
 twistor
superfields $f^I$ and $g_I$. Instead of simply making this
comparison on a term by term basis, it is more illuminating to
recognize that the chiral superfield ${\cal W}(x,\t)$ that is the
basic variable of linearized conformal supergravity coincides with
the field of the same name \eqn\idew{{\cal
W}(x,\theta)=\int_{\Bbb{D}_{x,\theta}}g_IdZ^I} that we introduced
at the end of section 3 in order to give a spacetime
interpretation to the twistor fields. Here $\Bbb{D}_{x,\theta}$ is
defined via the twistor equations \eqn\kini{\left(\mu^{\dot
a},\psi^A\right)=\left(x^{a\dot
a}\lambda_a,\theta^{Aa}\lambda_a\right),} where now we take the
variables to be real.

We can more explicitly write \eqn\nidew{{\cal W}(x,\t)= \int d\l^a
\left(g_a(Z) + x_{a\ad} g^\ad (Z) +\t_a^A g_A(Z)\right).} Here we
evaluated $dZ^I$ using \eqn\bidew{ \left(d\l^a, d\mu^\ad,
d\psi^A\right) = \left(d\l^a, d\l_b x^{b\ad}, d\l_c \t^{c
A}\right),} which holds when the $Z^I$ are evaluated on
$\Bbb{D}_{x,\theta}$, or in other words are regarded as functions
of $\lambda^a$ for fixed $x$ and $\theta$ via the twistor
equations \kini. On the right hand side of \nidew, these
conditions on $Z$ are understood.

 The equation of motion
for ${\cal W}$, derived from the superspace action \linact,   is
\eqn\hino{ \e^{ABCD} D^a_C D_{Ea} D_{b D} D^b_F {\cal W} =0.} This
implies but is stronger than the constraint \reality. This
equation of motion is automatically satisfied by any superspace
function ${\cal W}$ that can be written in terms of a twistor
field $g_I$ via \idew\ or \nidew.
 To see this,
first note that the equation of motion for ${\cal W}$ is
equivalent to $\triangle_{AB}\triangle_{CD}{\cal W}=0$, where
$\triangle_{AB}=D_{Aa}D_B^a$. (In writing the equation for ${\cal
W}$ this way, we appear to introduce extra equations absent in
\hino, but they are consequences of fermi statistics.) If
$\phi(Z)$ is any function of $Z$, we can convert it to a function
$\phi'(\lambda,x,\theta)$ by expressing $\mu$ and $\psi$ in terms
of $\lambda$, $x$, and $\theta$ via \kini. (In other words,
$\phi'(\lambda^a,x^{b\dot b},\theta^{Ac})=\phi(\lambda^a,x^{b\dot
b}\lambda_b,\theta^{Ac}\lambda_c)$.) For any function $\phi'$
obtained this way, it follows from the chain rule that
 $D_F^b\phi' = \l^b (\partial_F +\bar\t^\bd_F \p_\bd)\phi' $ where $\p_F =
{\p\over{\p\psi^F}}$ and $\p_\bd = {\p\over{\p\mu^\bd}}$.  From
this it follows, using the fact that $\lambda_a\lambda^a=0$, that
$\triangle_{AB}\phi'=0$. It does not follow from this that
$\triangle_{AB}{\cal W}=0$, because $\triangle_{AB}$ can act on
the explicit factors of $x$ and $\theta$ present on the right hand
side of \nidew.  However, since those factors are linear in $x$
and $\theta$, and $\triangle_{AB}$ annihilates
$[\triangle_{CD},x]$ and $[\triangle_{CD},\theta]$, it does follow
that $\triangle_{AB}\triangle_{CD}{\cal W}=0$.

It is more difficult to write the twistor fields $g_I$ in terms of
${\cal W}(x,\t)$.  As a step in this direction, first note that
\idew\ implies that \eqn\boxw{\p^\mu\p_\mu {\cal W} = \int d\l^a
(\p^\mu\p_\mu) x_{a\ad} g^\ad(\l,x\l,\t\l) = \int d\l^\a \l_a
\p_\ad g^\ad. } Using the description of the dipole fields in
section 4 and the component expansion of ${\cal W}$ in \compw, one
therefore finds that
\eqn\gaddef{ \p_\ad g^\ad(Z)=\hat C'_{0} + \psi^A
\hat\Lambda'_{-\half  A} + (\psi^2)_{[AB]} \hat T'_{-1}{}^{[AB]} +
(\psi^3)_A\hat \eta'{}^A_{-{3\over 2}} + \psi^4\hat e'_{-2},}
where $\hat C'_{0}$ denotes the twistor field of GL(1) charge $-2$
for the spacetime field $C'_{0}$ of helicity zero,
$\hat\Lambda'_{-\half  A}$ denotes the twistor field of GL(1)
charge $-3$ for the spacetime field $\Lambda'_{-\half  A}$ of
helicity $-\half$, etc.

To obtain $g^\ad$ from \gaddef,
one can use the usual
%NBnew  corrected and rephrased some statements in (4.25) - (4.30)
momentum-space description of twistor fields in which $g^\ad$ depends
on $\mu^\ad$ as $\d(\pi^a\l_a) \exp(i\mu^\ad
\bar\pi_\ad (\pi^1/\l^1))$,
where $k^{a\ad}
=\pi^a\bar\pi^\ad$ is the momentum and $(\pi^1/\l^1)$ is Lorentz-covariant
because of the delta-function $\d(\pi^a\l_a)$ in $g^\ad$.
So $\p_\ad g^\ad = i(\pi^1/\l^1)\bar\pi^\ad g_\ad$ and one can use
the gauge invariance $\d g_I = \p_I\Omega$ to choose the gauge
\eqn\gadfin{g_\ad(Z)= -i{{\l^a\s^0_{a\ad}}\over {k^0}} (\hat
C'_{0} + \psi^A \hat \Lambda'_{-\half  A} + (\psi^2)_{[AB]} \hat
T'_{-1}{}^{[AB]} + (\psi^3)_A\hat \eta'{}^{A}_{-{3\over 2}} + \psi^4\hat
 e'_{-2}).}
Although this gauge choice is not Lorentz-covariant since it singles out
the time direction, it is convenient
for comparing with the dipole solution of \csol\ which also singles out
the time direction.

To determine $g_A$, note that \idew\ implies that
\eqn\derivw{\s^{\mu}_{\a\ad} {\p\over{\p\t_a^A}}\p_\mu {\cal W} =
\int d\l^a \l_a \left(\p_A g_\ad(\l,x\l,\t\l) -\p_\ad
g_A(\l,x\l,\t\l)\right).} Comparing with \compw\ and \lambdasol\
and using that
$\p_\ad g_A = i(\pi^1/\l^1)\bar\pi_\ad g_A$, one finds that
\eqn\gAdef{\eqalign{g_A(Z)=&\hat\Lambda_{\half A} + \psi^B\hat
T_{0[AB]} +\psi^B\hat E_{0(AB)} +
(\psi^2)_{[AB]}\hat\eta_{-\half}^B  \cr &+
(\psi^2)_{[BC]}\hat\xi_{-\half A}^{[BC]} +(\psi^3)_A \hat e_{-1}
+(\psi^3)_B \hat V_{-1 A}^B +\psi^4\hat{\bar\eta}_{-{3\over 2}
A}.\cr}}
Note that this identification for $g_A(Z)$ is consistent with the
equation
\eqn\boxtw{\e_{ab}{\p\over{\p\t_a^A}}
{\p\over{\p\t_b^B}}
{\cal W}
 = \int d\l^\a \l_a
(\p_A g_B(\l,x\l,\t\l) +\p_B g_A (\l,x\l,\t\l)), } which follows from
\idew.

 Finally, one can use \idew\ to relate $g_a$ with ${\cal
W}$ by defining
$$ \int d\l^a g_a = {\cal W} -
\int d\l^a [x_{a\ad} g^\ad
+\t_a^A g_A],$$
which implies that
\eqn\gadef{g_a(Z)=\l_a(\hat C_0 + \psi^A\hat
\Lambda_{-\half A} + (\psi^2)_{[AB]}
\hat T_{-1}^{[AB]} + (\psi^3)_A \hat \eta^A_{-{3\over 2}} + \psi^4\hat
 e_{-2}) + ...}
%$$
%-{{\mu^\ad \s^0_{a\ad}}\over k^0} (\hat C'_{0} + \psi^A \hat
%\Lambda_{-\half ' A} + (\psi^2)_{[AB]} \hat T_{-1'}^{[AB]} +
%(\psi^3)_A \hat \eta^A_{-{3\over 2}'} + \psi^4\hat e_{-2'}).$$
where $...$ depends on fields appearing in $g_\ad$ and $g_A$.

Similarly, one can relate the antichiral superfield $\bar {\cal
W}(x+\t\bar\t,\bar \t)$ with the dual field  $\widetilde
g^I(\bar Z)$ of \fourier\ where $\bar Z_I$ plays the role of the dual
variable $V_I$ in \fourier. The identification is
\eqn\idetwo{\eqalign{\bar {\cal W}(\hat x,\bar\t) =& \int d\bar
Z_I \widetilde g^I(\bar Z|_{\bar\mu^\a = \hat x^{a\ad} \bar\l_a,
\bar\psi_A = \bar\t_A^\ad\bar\l_\ad})\cr =& \int d\bar\l^\ad
\left(\widetilde g_\ad(\bar\l,\bar\mu=\hat x\bar\l,\bar\psi=
\bar\t\bar\l) + \hat x_{a\ad} \widetilde g^a (\bar\l,\bar\mu=\hat
x\bar\l,\bar\psi= \bar\t\bar\l) \right.
\cr & +\left. \bar\t_{\ad A} \widetilde
g^A(\bar\l,\bar\mu=\hat x\bar\l,\bar\psi=\bar\t\bar\l)\right)\cr}}
where $\hat x^{a\ad} = x^{a\ad} + \t^{aA} \bar\t^{\ad}_A$. Using
similar arguments to those above and defining $f^I$ in terms of
$\widetilde g^I$ using \fourier, \idetwo\ implies that one can
choose a gauge such that
\eqn\urky{\eqalign{\l^a f_a(Z) =&\hat e'_{2 }+ \psi^A
\hat{\bar\eta}'_{{3\over 2} A} + (\psi^2)_{[AB]} \hat{\bar
T}'_{1}{}^{[AB]} + (\psi^3)_A\hat {\bar\Lambda}'_{\half }{}^A +
\psi^4\hat {\bar C}'_{0 },\cr f^A(Z)=& \hat\eta_{3\over 2}^A +
\psi^B\hat V_{1A}^B + \psi^A\hat e_1 +
(\psi^2)^{[BC]}\hat{\bar\xi}^A_{\half [BC]}\cr & +
(\psi^2)^{[AB]}\hat{\bar\eta}_{\half B} +(\psi^3)_B\hat {\bar
E}^{(AB)}_0 +(\psi^3)_B \hat{\bar T}_0^{[AB]} +\psi^4\hat
{\bar\Lambda}_{-\half}^A,\cr  f_\ad(Z) =&\p_\ad(\hat e_2 + \psi^A
\hat{\bar\eta}_{{3\over 2} A} + (\psi^2)_{[AB]}\hat {\bar
T}_1^{[AB]} + (\psi^3)_A\hat {\bar\Lambda}_{\half}^A + \psi^4\hat
{\bar C}_0) + ...
%\cr &- {\s^0_{a\ad}\over k^0} \p^a( \hat e_{2 '}+
,\cr}}
%\cr &- {\s^0_{a\ad}\over k^0} \p^a( \hat e_{2 '}+
%\psi^A \hat{\bar\eta}_{{3\over 2}' A} + (\psi^2)_{[AB]}\hat{\bar
%T}_{1'}^{[AB]} +
% (\psi^3)_A\hat {\bar\Lambda}_{\half '}^A +
%\psi^4\hat {\bar C}_{0 '}),\cr}}
where $\p_a = {\p\over{\p\l^a}}$,
$\p_\ad = {\p\over{\p\mu^\ad}}$, and $...$ depends on fields appearing
in $f_a$ and $f^A$.

\newsec{Some Tree-Level Scattering Amplitudes}

In this section, we evaluate some tree-level scattering
amplitudes. First we consider three-point amplitudes and then MHV
amplitudes.  The computations are mainly done using the
open-string version of twistor-string theory, but in section 5.3,
we compare a few statements to analogous statements based on the
$B$-model of $\Bbb{CP}^{3|4}$.

\subsec{Three-Point Tree Amplitudes}

The three-point tree amplitudes are computed from the correlation
function \eqn\gorun{ \langle V_1 (z_1) ~ V_2(z_2) ~
V_3(z_3)\rangle} at degree zero and degree one. We identify the
open string worldsheet with the upper half of the complex
$z$-plane; open string vertex operators are inserted on the real
axis. We do not write the ghosts explicitly; as usual, they give a
factor $|z_1-z_2||z_2-z_3||z_3-z_1|$ that cancels a similar factor
that arises in evaluating \gorun.

As explained in section 2.1, the vertex operators, for gauge
bosons or for supergravitons of appropriate helicities, are
respectively \eqn\vertices{V_\phi = j^r \phi_r(Z),\quad V_f = Y_I
f^I(Z),\quad V_g = \p Z^I g_I(Z).}

We first examine the degree zero contribution to the three-point
functions.  The correlators of the currents are familiar:
\eqn\uncu{\bigl\langle j^r(z_1)j^s(z_2)j^t(z_3)\bigr\rangle =
{kf^{rst}\over (z_1-z_2)(z_2-z_3)(z_3-z_1)}.} Here $k$ is the
level of the current algebra, and $f^{rst}$ are the structure
constants.  What about the correlators of vertex operators
constructed from fields on $\Bbb{RP}^{3|4}$? For functions
$\phi_i(Z)$ on $\Bbb{RP}^{3|4}$, the degree zero correlator is
simply
\eqn\buncu{\left\langle\prod_{i=1}^3\phi_i(Z(z_i))\right\rangle
=\int_{\Bbb{RP}^{3|4}}d\Omega\,\prod_i\phi_i(Z),} where $d\Omega$
is the usual measure.  Let us now try to evaluate a degree zero
correlator containing a single $Y$ field.  The basic vertex
operator containing the $Y$ field is $f^I(Z)Y_I$, where as in
section 2.1, $f^I(Z)$ is a volume-preserving vector field on
$\Bbb{RP}^{3|4}$. Using $\langle
Y_I(z)Z^J(w)\rangle=\delta_I^J/(z-w)$, we get
\eqn\vuvulo{\left\langle
f^I(Z)Y_I(z)\prod_{j=1}^n\phi_j(w_j)\right\rangle=\int_{\Bbb{RP}^{3|4}}d\Omega
\sum_{j=1}^n{1\over z-w_j}f^I(Z){\partial \phi_j(Z)\over\partial
Z^I}\prod_{k\not=j} \phi_k(Z).} Let us verify the $SL(2,\Bbb{R})$
invariance of this formula.  Under \eqn\junnon{z\to
(az+b)/(cz+d),~~~w_i\to (aw_i+b)/(cw_i+d),} \vuvulo\ transforms to
\eqn\huvuo{\int_{\Bbb{RP}^{3|4}}d\Omega \sum_{j=1}^n {
(cz+d)(cw_j+d)}{1\over z-w_j}f^I(Z){\partial
\phi_j(Z)\over\partial Z^I}\prod_{k\not=j} \phi_k(Z).} We would
like the amplitude \vuvulo\ under $SL(2,\Bbb{R})$ to be simply
multiplied by $(cz+d)^2$, that is, to transform to
\eqn\uvou{{(cz+d)^2}\int_{\Bbb{RP}^{3|4}}d\Omega
\sum_{j=1}^n{1\over z-w_j}f^I(Z){\partial \phi_j(Z)\over\partial
Z^I}\prod_{k\not=j} \phi_k(Z),} reflecting the fact that the
operator $f^I(Z)Y_I$ has conformal dimension 1 and the other
operators  have conformal dimension 0. The difference is
\eqn\tuvuo{-{c(cz+d)}\int_{\Bbb{RP}^{3|4}}d\Omega
\sum_{j=1}^nf^I(Z){\partial \phi_j(Z)\over\partial
Z^I}\prod_{k\not=j} \phi_k(Z)}
$$=-{c (cz+d)}
\int_{\Bbb{RP}^{3|4}}d\Omega \,f^I(Z){\partial \over\partial
Z^I}\prod_{k} \phi_k(Z).$$
This vanishes upon integrating by parts
and using the fact that $f^I$ is volume-preserving,
$\partial_If^I=0$.

Correlators with several $Y$ fields are evaluated similarly by
summing over contractions. For example,
\eqn\homonf{\eqalign{&\left\langle
f_1^IY_I(z_1)f_2^JY_J(z_2)f_3^KY_K(z_3) \right\rangle =\cr&{1\over
(z_1-z_2)(z_2-z_3)(z_3-z_1)}\int_{\Bbb{RP}^{3|4}}\left({\partial\over
\partial Z^K}f_1^I{\partial\over\partial
Z^I}f_2^J{\partial\over\partial Z^J}f_3^K- {\partial\over
\partial Z^J}f_1^I{\partial\over\partial
Z^K}f_2^J{\partial\over\partial Z^I}f_3^K\right).\cr}} Degree zero
correlators containing a field $\partial Z$ vanish unless there is
a $Y$ field to contract it with and are evaluated using $\langle
Y_I(z)\partial Z^J(w)\rangle=\delta_I^J/(z-w)^2$.

Using these rules, the nonvanishing degree zero three-point
functions come from \eqn\dzero{ \langle V_\phi V_\phi
V_f\rangle,\quad \langle V_f V_f V_f\rangle,\quad \langle V_f V_f
V_g\rangle.} For instance, we evaluated the $\langle
V_fV_fV_f\rangle$ correlator in \homonf.

A perplexing point, which corresponds to facts noted in section
3.2 of \witten,\foot{See also a footnote on p. 7 of \berkmotl.} is
that, combining \uncu\ and \buncu, the correlator $\langle
V_{\phi_1}(z_1) V_{\phi_2}(z_2) V_{\phi}(z_3)\rangle$ is nonzero
but antisymmetric in $1,\,2$, and 3, as a result of which,
allowing for Bose statistics (and including the ghosts), the
three-point coupling vanishes after summing over different cyclic
orderings of $z_1$, $z_2$, and $z_3$.  This three-point function
would correspond to the three-gluon tree amplitude with helicities
$++\,-$ (and other amplitudes related to this by supersymmetry).
This amplitude is non-vanishing in Yang-Mills theory for generic
on-shell complex momenta, though it vanishes on-shell for real
momenta in Lorentz signature.  As explained in section 4.3 of
\witten, this amplitude is nonzero in the complex version of
twistor space.  (In that context, the fields carry an extra
antiholomorphic index, which avoids the problem with Bose
statistics.)

The above correlators lead to a variety of three-point amplitudes
for helicity states. Among others, these include three-point
amplitudes for states $A_{\pm 1}$ of the Yang-Mills gauge field
with helicities $\pm 1$, states $e_{\pm 2}$ of the graviton with
helicities $\pm 2$, and helicity zero states of the scalar fields
$C$ and $\bar C$. If we allow ourselves the liberty of including
the $\langle V_\phi V_\phi V_\phi\rangle$ coupling whose anomalous
status was noted in the last paragraph, then the correlators
identified above lead to the following couplings: \eqn\cubiczero{
A_1 A_1 A_{-1},\quad A_1 A_1 \bar C,\quad A_1 A_{-1} e_2,\quad
\bar C e_2 e_2,\quad e_2 e_2 e_{-2},\quad \bar C e_2 C.} To reach
this conclusion, we use the results of section 3 (and of \witten\
in the case of $V_\phi$). Vertex operators $V_\phi$, $V_f$, and
$V_g$ describe supermultiplets whose bottom and top components are
respectively $A_1+\dots+\psi^4A_{-1}$, $e_2+\dots+\psi^4\bar C$,
and $C+\dots +\psi^4e_{-2}$.  Evaluation of degree zero
correlators entails an integral over $\Bbb{RP}^{3|4}$, as seen in
all the formulas above; this picks out, among other things,
amplitudes with two bottom components and a top component.

Now we move to degree 1.  As explained in \witten, section 3.1, a
degree 1 curve is a ``line'' $\Bbb{D}_{x,\theta}$.  It has
homogeneous coordinates $\lambda^a$ and is described by the
familiar equations $\mu^{\dot a}=x^{a\dot a}\lambda_a$,
$\psi^A=\theta^{Aa}\lambda_a$, where $x^{a\dot a}$ and
$\theta^{Aa}$ are the moduli of the curve $\Bbb{D}_{x,\theta}$.
The string worldsheet, which we describe as the complex $z$-plane,
is mapped to $\Bbb{D}_{x,\theta}$ by $\lambda^1=az+b$,
$\lambda^2=cz+d$, where $a,b,c$, and $d$ are specified up to a
constant multiple. In computing a degree 1 contribution to a
scattering amplitude of open strings, we must as always impose a
gauge condition to fix the $SL(2,\Bbb{R})$ invariance of the open
string worldsheet. While this can be done by imposing the
condition that three of the open strings are inserted at specified
values of $z$, say $0,1$, and $\infty$, the instanton computation
is simplest if one imposes $SL(2,\Bbb{R})$ invariance by
parametrizing the instanton as $\lambda^1=1$, $\lambda^2=z$, and
integrates over all insertion positions for the vertex operators.
The degree 1 contribution to the scattering amplitude is thus
computed from the formula \eqn\onnnno{\int
d^4x\,d^8\theta\int_{\Bbb{D}_{x,\theta}} dz_1\dots dz_n\langle
V_1(z_1)\dots V_n(z_n)\rangle.}

Using this recipe, it is straightforward to identify the following
nonzero three-point functions: \eqn\dzero{\langle V_\phi V_\phi
V_\phi\rangle,\quad \langle V_\phi V_\phi V_g\rangle,\quad \langle
V_g V_g V_g\rangle,\quad \langle V_g V_g V_f\rangle.}  Very few
contractions of $\Bbb{RP}^{3|4}$ fields (as opposed to currents of
the current algebra) are needed to evaluate these correlators; in
fact, the only such contraction  is a $\langle Y\partial Z\rangle$
contraction that is needed to evaluate $\langle V_gV_gV_f\rangle$.
For example, $\langle V_gV_gV_g\rangle$ can be evaluated in degree
1 without any quantum contraction at all; it just equals $\int
d^4x d^8\theta
\left(\int_{\Bbb{D}_{x,\theta}}g_IdZ^I\right)^3=\int
d^4x\,d^8\theta\,{\cal W}^3$, a formula that was essentially
explained in section 2.2.

Concentrating on the same helicities as before, these amplitudes
describe the cubic couplings
\eqn\nnonno{A_{-1}A_{-1}A_1,~~A_{-1}A_{-1}C,~~Ce_{-2}e_{-2},~~e_{-2}e_{-2}e_2,~~
Ce_{-2}\bar C.} These results arise because the degree one curve
$\Bbb{D}_{x,\theta}$ has eight fermionic moduli $\theta^{Aa}$;
integration over them generates amplitudes with two top components
in the supermultiplets.   The amplitudes in \nnonno\ are parity
conjugates of the couplings that appeared at degree zero.

The cubic couplings we have obtained are consistent with the
action \eqn\cubicaction{ \eqalign{S=\int d^4 x &\biggl(
\exp(2\bar C) \bigl(F_{\ad\bd} F^{\ad\bd} + W_{\ad\bd\dot c\dot
d} W^{\ad\bd\dot c\dot d} + \square^2  C\bigr)\biggr. \cr
&+\biggl.\exp(2C) \bigl(F_{ab} F^{ab} + W_{abcd} W^{abcd} +
\square^2 \bar C\bigr)\biggr).\cr}} This will be qualitatively
%NBnew sign of C coupling?
compared in section 6 to expectations from conformal supergravity.
(We have here included the kinetic energy of the $C$ field in the
form suggested by the discussion in section 6, though, as it leads
to vanishing on-shell $CC\bar C$ and $C\bar C\,\bar C$ couplings,
it is not detected by the above computation.)

\subsec{MHV Tree Amplitudes}

In this section, by evaluating \onnnno, we will compute MHV tree
amplitudes that include supergravitons in addition to possible
gauge bosons. We will not compute all such amplitudes, but we will
compute a set of them that contains amplitudes for gluons and for
gravitons of either helicity as well as for the dilatonic fields
$C$ and $\bar C$.

We recall from section 2.1 that supergravitons are described by
vertex operators $V_f=f^I(Z)Y_I$ and $V_g=g_I(Z)dZ^I$, where $f^I$
is a volume-preserving vector field on $\Bbb{RP}^{3|4}$, and $g_I$
is an abelian gauge field on $\Bbb{RP}^{3|4}$.  To compute
amplitudes with external gravitons and dilatons, it suffices, as
we know from our analysis of the spectrum in section 3, to
consider the components $f^a, \,f^{\dot a}$ of $f^I$ (as opposed
to $f^A$, which describes other modes), and similarly it suffices
to consider the components $g_a, \,g_{\dot a}$ of $g_I$ (as
opposed to $g_A$).

In addition, to keep things simple, we will take external
supergravitons to have wavefunctions that are plane waves
$\exp(ik\cdot x)$, as opposed to the more general wavefunctions
$A\cdot x \,\exp(ik\cdot x)$ encountered in section 4.  Plane
waves are wavefunctions on which the translations can be
diagonalized.  As we noted at the end of section 3, the vertex
operators with this property are $f^{\dot a}Y_{\dot a}$ and $g_a
\partial Z^a$.

Twistor space wavefunctions that correspond to plane waves in
Minkowski spacetime have been described most fully (for gluons) in
section 2.1 of \ref\uwitten{E. Witten, ``Parity Invariance For
Strings In Twistor Space,'' hep-th/0403199.}.  Before reviewing
these wavefunctions in the next paragraph, we recall some standard
conventions. If $\alpha$ and $\beta$ are two positive chirality
spinors, we write $\langle\alpha,\beta\rangle$ as an abbreviation
for $\epsilon_{ab}\alpha^a\beta^b$.  Similarly, if
$\tilde\alpha^{\dot a}$ and $\tilde\beta^{\dot b}$ have negative
chirality, we write $[\tilde\alpha,\tilde\beta]$ for
$\epsilon_{\dot a\dot b}\tilde\alpha^{\dot a}\tilde\beta^{\dot
b}$.  Given massless particles of momenta $p^i_{a\dot
a}=\pi^i_a\tilde\pi^i_{\dot a}$, we also write $\langle
i,j\rangle$ for $\langle\pi_i,\pi_j\rangle$, and $[i,j]$ for $[
\tilde\pi_i,\tilde\pi_j]$.

The twistor space wavefunction of a massless Yang-Mills particle
with definite momentum $p_{a\dot a}=\pi_a\tilde\pi_{\dot a}$ is
$V_\phi=\sum_r\phi_r(\lambda,\mu,\psi) \cdot j^r$, where $j^r$ are
the currents, and roughly speaking each $\phi^r$ is a multiple of
\foot{See \refs{\witten,\uwitten} for more information on why this
type of wavefunction describes a plane wave in spacetime.  Roughly
speaking, it describes a state of definite $\pi$ because of the
delta function, and a state of definite $\tilde\pi$ because of the
exponential dependence on $\mu$, so it describes a state of
definite momentum $p_{a\dot a}=\pi_a\tilde\pi_{\dot a}$.}
\eqn\unonn{\phi(\lambda,\mu,\psi)=\delta(\langle\lambda,\pi\rangle)
\exp(i[\mu,\tilde\pi]) u(\psi).} Here the delta function has
support on the locus where (up to scaling) $\lambda^a=\pi^a$, and
the fermionic wavefunction $u(\psi)$ determines which helicity
state we get in the Yang-Mills multiplet.  Actually, \unonn\ needs
to be corrected slightly to get the right homogeneity in all
variables. Since on the support of the delta function, $\lambda$
is a multiple of $\pi$, there is a well-defined ratio
$(\pi/\lambda)$. The refined version of \unonn\ that we really
want is
\eqn\bunonn{\phi(\lambda,\mu,\psi)=(\lambda/\pi)\delta(\langle\lambda,\pi\rangle)\cdot
\exp(i[\mu,\tilde\pi](\pi/\lambda))\cdot u((\pi/\lambda)\psi).}
The factors of $(\pi/\lambda)$ make everything scale properly. The
wavefunction is homogeneous in twistor coordinates
$Z^I=(\lambda,\mu,\psi)$ of degree zero (this is the right scaling
for $\phi$, as we recall from section 2.1).

Plane wave states of supergravitons with momentum $p_{a\dot
a}=\pi_a\tilde\pi_{\dot a}$ can be described by analogous twistor
space wavefunctions. One type of vertex operator is $V_f=f^{\dot
a}Y_{\dot a}$, where \eqn\nconco{f^{\dot
a}(\lambda,\mu,\psi)=\tilde\pi^{\dot a}
(\lambda/\pi)^2\delta(\langle\lambda,\pi\rangle)\cdot
\exp(i[\mu,\tilde\pi](\pi/\lambda)) \cdot u((\pi/\lambda)\psi).}
We take $f^{\dot a}\sim \tilde\pi^{\dot a}$ in order to satisfy
the volume-preserving condition $\partial f^{\dot a}/\partial
\mu^{\dot a}=0$.  The factors of $(\pi/\lambda)$ ensure that
$f^{\dot a}$ is homogeneous in twistor coordinates of weight one.
The other vertex operator we need is $V_g=g_a
\partial Z^a$,
where
\eqn\looly{g_a(\lambda,\mu,\psi)=\lambda_a(\pi/\lambda)\delta(\langle
\lambda,\pi\rangle)\cdot \exp(i[\mu,\tilde\pi](\pi/\lambda))\cdot
u((\pi/\lambda)\psi).}  In this case, we take $g_a\sim \lambda_a$
to obey the constraint $\lambda^ag_a=0$.  The factors of
$(\pi/\lambda)$ ensures that $g_a$ is homogeneous in twistor
variables of degree $-1$.

 Under $(\pi,\tilde\pi)\to (t\pi,t^{-1}\tilde\pi)$,
the $\psi=0$ components of $V_\phi$, $V_f$, and $V_g$ scale as
$t^{-2}$, $t^{-4}$, and $1$, reflecting the fact that those
operators describe states in Minkowski spacetime of helicities
$1,2$, and 0. In general, a state of helicity $h$ is represented
by a vertex operator that scales as $t^{-2h}$.  The scattering
amplitudes obtained by evaluating the expectation value of the
product of vertex operators will likewise scale as $t^{-2h}$.

In a multi-particle scattering amplitude, let $\Phi$ be the set of
external particles with vertex operators of type $V_\phi$, $F$ the
set of external particles with vertex operators $V_f$, and $G$ the
set of external particles with vertex operators $V_g$. Let
$N_\Phi$, $N_F$, and $N_G$ be the number of elements of $\Phi$,
$F$, and $G$, and let $N=N_\Phi+N_F+N_G$ be the total number of
external particles. Let $p_k^{a\dot a}=\pi_k^a\tilde\pi_k^{\dot
a}$ be the momentum of the $k^{th}$ particle, let $z_k$ be
worldsheet coordinate at which it is asserted, and let
$(\lambda_k,\mu_k,\psi_k)=(\lambda(z_k),\mu(z_k),\psi(z_k))$
parametrize the image of $z_k$ in twistor space.  Finally, let
$u_k((\pi_k/\lambda_k)\psi_k)$ be the fermionic wavefunction of
the $k^{th}$ particle. In these definitions, $k$ runs over all $N$
possible values.

Now we commence to evaluate the scattering amplitudes. To evaluate
the $d^4x$ integral in \onnnno, we use the fact that on the curve
$\Bbb{D}_{x,\theta}$, $\mu^{\dot a}=x^{a\dot a}\lambda_a$. $\mu$
enters our wavefunctions only via the exponential factors, and
therefore in any product of the above-described vertex operators
of particles of momenta $p_j^{a\dot a}=\pi_j^a\tilde\pi_j^{\dot
a}$, the $x$-dependence is of the form
\eqn\xdepe{\exp\left(ix_{a\dot a}\sum_j\pi_j^a\tilde\pi_j^{\dot
a}\right).} We have used the delta functions in the wavefunctions
to set $\lambda^a$ of the $j^{th}$ particle to a multiple of
$\pi_j$; the multiple conveniently cancels out because of the
$(\pi/\lambda)$ factor in the exponent in \bunonn. This is a
typical example of how those factors often disappear in
calculations, by helping turn $\lambda$'s into $\pi$'s. The $x$
integral in \onnnno\ therefore gives simply a delta function of
energy-momentum conservation, $(2\pi)^4\delta^4(\sum_jp_j^{a\dot
a})$.

We use the scaling symmetry of the homogeneous coordinates
$(\lambda,\mu,\psi)$ of $\Bbb{CP}^{3|4}$ to set $\lambda^1=1$.
With  the open string worldsheet understood as the upper half
$z$-plane, a degree one instanton has $\lambda^2=(az+b)/(cz+d)$
for some real $a,b,c$, and $d$; as discussed in obtaining \onnnno,
we fix the $SL(2,\Bbb{R})$ invariance so that $\lambda^2=z$. It
follows, for example,  that the factor $\lambda_a
\partial \lambda^a$  in the vertex
operators of type $V_g$ is equal to 1.  Indeed, $\partial$ is just
$\partial/\partial z$, so with $(\lambda^1,\lambda^2)=(1,z)$, we
get $(\partial\lambda^1,\partial\lambda^2)=(0,1)$. Moreover,
\eqn\kinon{\lambda_k/\pi_k=1/\pi_k^1,~~z_j-z_k=-\langle
\pi_j,\pi_k\rangle/\pi_j^1\pi_k^1.}

 The vertex operator for the $j^{th}$ external particle
  contains a delta function $\delta(\langle\lambda,\pi_j\rangle)$,
which now becomes $\delta(\pi_j^2-z_j\pi^1_j)$.  The $z_j$
integrals can be done with the help of these delta functions, with
with the result that the $j^{th}$ particle is inserted at
$z_j=\pi_j^2/\pi_j^1$, and the amplitude acquires a factor
\eqn\honno{\int
dz_j\delta(\pi_j^2-z_j\pi_j^1)=1/\pi_j^1=(\lambda_j/\pi_j).}

Let $\pi_\Phi$ and $z_\Phi$ denote the collection of variables
$\pi_i$ and $z_i$ for $i\in \Phi$. In evaluating the scattering
amplitude, we must evaluate the current correlation function
${\cal J}_0(z_\Phi)=\langle\prod_{k\in \Phi}J^{r_k}(z_k)\rangle$.
Here by translation invariance, ${\cal J}_0(z_\Phi)$ is a function
only of the differences $z_i-z_j$, which are written in terms of
$\pi_k$ and $\lambda_k$ in \kinon, and of the Lie algebra indices
$r_k$. $SL(2,\Bbb{R})$ invariance implies, since the currents
$J^r$ have dimension 1, that ${\cal J}_0(z_\Phi)$ can be written
\eqn\orgo{{\cal J}_0(z_\Phi)={\cal J}(\pi_\Phi)\prod_{i\in
\Phi}(\pi_i/\lambda_i)^2} (times a group theory factor which we
suppress) for some function ${\cal J}(\pi_\Phi)$ that is
homogeneous of degree $-2$ in each $\pi_i$. For example, the most
familiar case is the case that the gauge group $G$ is a unitary
group, and after arranging the particles in $\Phi$ in a definite
cyclic order, say $1,2,\dots,N_\Phi$, we extract a single-trace
amplitude. In this case, \eqn\undon{{\cal J}(\pi_\Phi)=\prod_{i\in
\Phi}{1\over \langle i,i+1\rangle}.} This is a familiar factor in
MHV scattering amplitudes for gluons, first interpreted as a
current correlation function by Nair \ref\nair{V. P. Nair, ``A
Current Algebra For Some Gauge Theory Amplitudes,'' Phys. Lett.
{\bf B214} (1988) 215.}.

For every $j\in F$, the corresponding vertex operator contains a
factor $(\lambda_j/\pi_j)^3\tilde\pi_j^{\dot a}Y_{\dot a}$. The
factor of $(\lambda_j/\pi_j)^3$ is included here purely for
convenience.  After extracting it, the rest of the vertex operator
$V_f$ for this particle is proportional to $(\pi_j/\lambda_j)$;
this factor cancels the factor of $(\lambda_j/\pi_j)$ that comes from the $z_j$
integral in \honno.  For vertex operators $V_g$, a similar cancellation occurs
more directly; the vertex operator for $j\in G$
is proportional to $(\pi_j/\lambda_j)$, which cancels the factor coming from
the $z_j$ integral.  Vertex operators $V_\phi$ are proportional instead
to $(\lambda_j/\pi_j)$, but in this case the current correlation function gives
a factor of $(\pi_j/\lambda_j)^2$, as in \orgo; these factors combine to
$(\pi_j/\lambda_j)$, which again cancels the factor coming coming from the $z_j$
integral.

Returning to the factor $(\lambda_j/\pi_j)^3\tilde\pi_j^{\dot
a}Y_{\dot a}$ in a vertex operator of type $V_f$, it
 must be evaluated using the contraction $\langle Y_{\dot
a}(z) \mu^{\dot b}(z')\rangle =\delta^{\dot b}_{\dot a}/(z-z')$.
The only $\mu$'s in the wavefunctions are in the exponentials, and
upon evaluating the contraction, we get a factor
\eqn\nonnb{i(\lambda_j/\pi_j)^3\sum_{k\not=
j}{[j,k](\pi_k/\lambda_k)\over z_j-z_k}.} The sum over $k$
includes particles of all types (from Yang-Mills or gravity
multiplets) with $k\not= j$. With the aid of \kinon, we rewrite
\nonnb\ in the form \eqn\nonnbo{-i\sum_{k\not= j}{[j,k]\over
\langle j,k\rangle}{(\pi_k^1)^2\over (\pi_j^1)^2}.} If we
introduce a spinor $\zeta^a$ with components $\zeta^a=(0,1)$, we
can write this factor as \eqn\jicon{-i\sum_{k\not= j}{[j,k]\langle
k,\zeta\rangle^2\over
                      \langle j,k\rangle \langle j,\zeta\rangle^2}.}
This formula has the amazing property of being independent of
$\zeta$ as long as $\zeta\not= \pi_j$ (where it is ill-defined
because the denominator vanishes), and therefore it is covariant
though the intermediate steps in the derivation were not
manifestly covariant. To see the $\zeta$-independence, note that
any change in $\zeta$ takes the form $\zeta\to v\zeta + w \pi_j$
for some scalars $v$ and $w$. As \jicon\ is homogeneous in $\zeta$
of degree zero, we can set $v=1$.  Under $\zeta\to \zeta+w\pi_j$,
the denominator of \jicon\ is invariant, as $\langle
j,j\rangle=0$. The numerator is also invariant, after summing over
$k$, because by momentum conservation \eqn\onnnsp{\sum_{k\not=
j}[j,k]\langle k,\zeta\rangle =\sum_{k\not= j}[j,k]\langle
k,j\rangle=0.} (Momentum conservation states that $\sum_k
|k]\langle k|=0$; in the sums in \onnnsp,  the restriction to
$k\not= j$ is immaterial since $[j,j]=0$.) We can reduce \jicon\
to a covariant formula by setting $\zeta$ to equal one of the
$\pi_k$, and having done so, we can restore bose symmetry by
averaging over choices of $\zeta$. But it seems more illuminating
to leave the expression in the form given here.

When we combine all this, we learn that the tree level MHV
scattering amplitude for these fields is \eqn\bonnon{(-i)^F{\cal
J}(\pi_\Phi)\prod_{j\in F}\sum_{k\not= j}{[j,k] \langle
k,\zeta\rangle^2\over \langle j,k\rangle \langle j,\zeta\rangle^2}
\int d^8\theta^{Aa} \prod_{m=1}^Nu_m((\pi_m/\lambda_m)\psi_m).} We
recall that $u_m$ is the fermionic part of the wavefunction of the
$ m^{th}$ particle. This wavefunction is to be evaluated on the
instanton configuration, that is, for
$\psi_m^A=\theta^{Aa}\lambda_{m\,a}$.

The integral over the fermionic parameters $\theta^{Aa}$ of the
instanton that appears here is familiar from evaluating MHV tree
amplitudes of Yang-Mills theory. (That is the special case that
all vertex operators are of type $V_\phi$; the fermionic integral
in \bonnon\ does not distinguish the different types of vertex
operator.)  Consider the illuminating special case that $u_m=1$
for all values of $m$ except two, while $u_m={1\over
4!}\epsilon_{ABCD}\psi_m^A\psi_m^B\psi_m^C\psi_m^D(\pi_m/\lambda_m)^4$
for the remaining two cases, say $m=r$ and $s$.  This means that
all particles other than particles $r$ and $s$
 have the maximum helicity
in their multiplets,    while particles $r,s$ have the minimum
helicity in their multiplets.  The maximum helicity is 1, 2, or 0
for particles described by vertex operators $V_\phi$, $V_f$, or
$V_g$, and the minimum helicities are $-1$, 0, or $-2$.

In this type of
 example, the integral over the $\theta^{Aa}$ gives a factor of $\langle
r,s\rangle^4$, which is a familiar factor in the MHV tree
amplitudes of Yang-Mills theory.  So at last, the scattering
amplitude becomes \eqn\bonnon{(-i)^F{\cal J}(\pi_\Phi)\langle
r,s\rangle^4\prod_{j\in F}\sum_{k\not= j}{[j,k] \langle
k,\zeta\rangle^2\over \langle j,k\rangle \langle
j,\zeta\rangle^2}.} This formula describes the MHV tree amplitude
for scattering of $N-2$ particles which are either gauge bosons of
helicity 1, gravitons of helicity 2, or dilatonic scalars $C$, and
two particles, labeled $r$ and $s$, which are either gauge bosons
of helicity $-1$, scalars $\bar C$, or gravitons of helicity $-2$.

While MHV tree level amplitudes for gauge boson scattering are
functions only of $\pi $ and not $\tilde\pi$, we see from \bonnon\
that MHV tree level amplitudes with supergravitons have a
non-trivial but polynomial dependence on $\tilde\pi$.  In the
terminology of \witten, this corresponds to scattering amplitudes
with ``derivative of a delta function'' support on curves of
degree one.

\subsec{Comparison To $B$-Model Of $\Bbb{CP}^{3|4}$}

Finally, we will briefly attempt to interpret the results of
section 5.1 on cubic tree level couplings of supergravitons in
terms of the alternative approach to twistor-string theory via the
$B$-model of $\Bbb{CP}^{3|4}$.

We recall from section 2.2 that vertex operators $V_f=f^IY_I$
correspond in complex twistor space to disturbances in the almost
complex structure $J$.  On the other hand, vertex operators
$V_g=g_I\partial Z^I$ correspond to disturbances in the $B$-field
$b$.  In section 2.2, we described a tree level coupling $\int
b\wedge N(J) \,\Omega$,  where $N(J)$ is the Nijenhuis tensor.  As
$N(J)$ is a nonlinear function of $J$, this includes a $bJ^2$
coupling. That $bJ^2$ coupling is the complex twistor space analog
of the degree zero $f^2g$ coupling that we found in section 5.1.

In section 5.1, however, we also found a degree zero $f^3$
coupling.  To what in complex twistor space does this correspond?
It should correspond to a local coupling on $\Bbb{CP}^{3|4}$
(local because of the degree zero property) that contains a term
nonlinear in $J$ but independent of $b$.

A natural candidate for this term is the integral of a certain
Chern-Simons $(0,3)$-form that we will describe presently.  Once
this form, which we will call $\omega_{CS}(J)$, is constructed,
the interaction we want is
\eqn\nxxno{\int_{\Bbb{CP}^{3|4}}\omega_{CS}(J)\, \Omega.} It is
independent of $b$ and  nonlinear in $J$.  When expanded around
the standard $\Bbb{CP}^{3|4}$, it leads to the desired $J^3$
interaction.

To construct this Chern-Simons form, we first note that on any
manifold, there is a bundle $T^*$ of one-forms.  On an almost
complex manifold, we have a decomposition
\eqn\kinont{T^*=T^*_{1,0}\oplus T^*_{0,1}}
 of $T^*$ into the
bundles of forms of type $(1,0)$ and $(0,1)$, respectively.   The
$\partial$ and $\bar\partial$ operators on zero-forms (or
functions) are defined as follows:  if $f$ is a function, then
define $\bar\partial f$ and $\partial f$ by writing $df=\partial
f\oplus \bar\partial f$, where $\partial f$ and $\bar\partial f$
are of type $(1,0)$ and $(0,1)$, respectively. No integrability of
$J$ is required in any of these statements.

Similarly, the bundle of two-forms can be decomposed as the direct
sum of bundles of forms of type $(2,0)$, $(1,1)$, and $(0,2)$.  We
let $\pi_{1,1}$ be the projection operator from all two-forms to
forms of type $(1,1)$.

 Now
we want to define a $\bar\partial$ operator on $T^*_{1,0}$. This
operator, which we will call $\bar D$, will  map  $(1,0)$-forms to
$(1,1)$-forms.  We define it to map a $(1,0)$-form $\lambda$ to
\eqn\nncon{\bar D\lambda=\pi_{1,1}(d\lambda).} Suppose that $f$ is
a function. Then $d(f\lambda)=f\,d\lambda+df\wedge\lambda. $  The
above definition of $\bar D$ leads to \eqn\nncinn{\bar
D(f\lambda)=\bar\partial f\wedge\lambda+f\,\bar D\lambda.} This is
the defining property of a connnection, or in this case, of the
$(0,1)$ part of a connection.  It means that locally
\eqn\nnxnn{\bar D=d\bar X^{\bar I}\left({\partial\over
\partial\bar X^{\bar I}}+\alpha_{\bar I}\right)} for some
$\alpha_{\bar I}$.  $\bar D$ was defined to act on $T^*_{1,0}$, so
the components of $\alpha_{\bar I}$ are matrices acting on
$T^*_{1,0}$ (that is, they are endomorphisms of $T^*_{1,0}$).
Thus, $\alpha_{\bar I}$ is a ``gauge field,'' or at least the
$(0,1)$ part of one.  So one can construct from $\alpha_{\bar I}$
a Chern-Simons $(0,3)$-form in the standard fashion:
\eqn\gloomy{\omega_{CS}(J)=d\bar X^{\bar I}d\bar X^{\bar J}d\bar
X^{\bar K}\,\,\Tr_{T^*_{1,0}}\left(\alpha_{\bar I}\partial_{\bar
J}\alpha_{\bar K} +{2\over 3}\alpha_{\bar I}\alpha_{\bar
J}\alpha_{\bar K}\right).} This $(0,3)$-form is then used in
\nxxno\ to construct the desired interaction.

\newsec{Conformal Supergravity Action}

As we explained at the end of section 2, the contribution of an
instanton of degree $d$ to a scattering amplitude is proportional
to $\exp(-d\langle C\rangle)$, where $C$ is the lowest component
of the superfield ${\cal W}$, and $\langle C\rangle$ is its
expectation value.

Consider now an $L$-loop twistor-string amplitude for the
scattering of $N$ external gluons and gravitons. (Even when there
are no external gravitons, this amplitude will not necessarily
coincide with a similar amplitude in supersymmetric Yang-Mills
theory, because supergravitons will appear as intermediate
states.) According to \witten, degree $d$ curves of genus $L$
contribute to $N$-gluon scattering processes with precisely
$d+1-L$ gluons of negative helicity, and therefore $N-1-d+L$
positive helicity gluons.\foot{We consider only connected
instantons; it has become reasonably clear \ref\roiban{R. Roiban,
M. Spradlin, and A. Volovich, ``On The Tree Level $S$ Matrix Of
Yang-Mills Theory,'' hep-th/0403190. } \nref\mhvtree{F. Cachazo,
P. Svrcek, and E. Witten, ``MHV Vertices And Tree Amplitudes
In Gauge Theory,''  hep-th/0403047. }%
\nref\gukov{S. Gukov, L. Motl, and A. Neitzke, ``Equivalence Of Twistor
Prescriptions For Super Yang-Mills,'' hep-th/0404085. }%
that the full connected twistor-string amplitudes can be computed
just from these contributions, a fact which is also manifest in
the open-string approach \berk\ since in that approach there are
no disconnected instantons. It has also become fairly clear
\refs{\mhvtree,\gukov} that the same amplitudes can be computed
from totally disconnected instantons.} Like the positive and
negative helicity gluon, the positive and negative helicity
graviton appears in the twistor fields at order $\psi^0$ and
$\psi^4$. The selection rules for scattering amplitudes with
positive and negative helicity particles are therefore the same
whether the particles are gluons or gravitons. So $L$-loop
amplitudes involving $N_-$ negative-helicity gluons and gravitons
depend on $\langle C\rangle$ as $\exp(-(N_- + L-1)\langle
C\rangle)$. By parity symmetry, this implies that amplitudes
involving $N_+$ positive-helicity gluons and gravitons depend on
$\langle\bar C\rangle$ as $\exp\left(-(N_+ +L-1)\langle\bar
C\rangle\right)$. (Of course, this result, like parity symmetry
itself, is much less obvious in the twistor formalism.) Combining
these results implies that $L$-loop scattering amplitudes for $N$
gluons and gravitons depend on $\langle C\rangle$ and $\langle\bar
C\rangle$ as \eqn\depende{ \exp\left( -(2L+N-2)\left\langle{C+\bar
C \over 2}\right\rangle + h\left\langle {C-\bar C\over
2}\right\rangle\right)} where $h = N_+ - N_-$.

The $L$-loop amplitude for scattering of $N$ gluons and gravitons further
depends on the string coupling constant $g_s$ as $(g_s)^{2L+N-2}$.
A look at \depende\ shows that $g_s$ can be absorbed into the
expectation value of $C$ and $\bar C$ by shifting $C\to C
+\log(g_s)$ and $\bar C\to\bar C +\log(g_s)$.

What kind of conformal supergravity action would be consistent
with this behavior?  As we have exploited in section 4.3, the
linearized ${\cal N}=4$ conformal supergravity can be described in
terms of a chiral superfield ${\cal W}$ which is posited to obey
the constraint \reality.  To get to the nonlinear level, one
should introduce some suitable potentials from which a suitable
nonlinear version of ${\cal W}$ can be constructed, in such a way
that a constraint generalizing \reality\ emerges as a Bianchi
identity. To our knowledge, this has not been done. If it can be
done, and a suitable chiral superspace measure $E(x,\t)$ can be
constructed from the underlying potentials (and is invariant under
constant shifts in $C$), then a supergravity action with the
property we want might take the form
 \eqn\classa{ {\cal S} = \int
%NBnew  sign of W?
d^4 x \int d^8\t \,E(x,\t) e^{2{\cal W}(x,\t)} + \int d^4 x \int
d^8\bar\t \,\bar E(\hat x,\bar\t) e^{2\bar {\cal W}(\hat
x,\bar\t)}.}  If the linearized theory is a good guide, ${\cal W}$
has zero conformal weight, so assuming the existence of a chiral
superspace measure, the action \eqn\supers{{\cal S}=\int d^4 x
\int d^8\t\, E(x,\t) f({\cal W}) + \int d^4 x \int d^8\bar\t
\,\bar E(x,\bar\t) \bar f(\bar {\cal W}) } is supersymmetric for
any holomorphic function $f({\cal W})$. This is analogous to the
case of ${\cal N}=2$ Yang-Mills theory, where for any
gauge-invariant function $f_{YM}({\cal W}_{YM})$ of the
adjoint-valued chiral superfield ${\cal W}_{YM}$, the action
\eqn\upers{\int d^4xd^4\theta \,f_{YM}({\cal W}_{YM})+\int
d^4xd^4\bar\t\, \bar f_{YM}(\bar {\cal W}_{YM})} is
supersymmetric.

In the ``minimal'' version of ${\cal N}=4$ conformal supergravity,
which was first proposed in \berg, the classical action was
assumed to be invariant under a global $SL(2,\Bbb{R})$ symmetry,
which includes invariance under constant shifts of $C$. This would
imply that $f({\cal W})={\cal W}^2$.  This choice is indeed
analogous to the choice most often made in ${\cal N}=2$ super
Yang-Mills theory, where the minimal theory (which in fact is
renormalizable) arises if $f_{YM}$ is quadratic.\foot{  As briefly
discussed in footnote {\it (c)} of \ref\tseytlin{A. Tseytlin, ``On
Limits Of Superstring In ${\rm AdS}_5\times\Bbb{S}^5$,'' Theor.
Math. Phys. {\bf 133} (2002) 1376, hep-th/0101112.}, one way to
construct a minimal action for ${\cal N}=4$ conformal supergravity
is as follows. Couple ${\cal N}=4$ super Yang-Mills theory to
background fields of conformal supergravity; compute the one-loop
effective action for these background fields and extract its Weyl
anomaly (that is, the change in the effective action under a
global rescaling of the background metric \ref\duff{D. M. Capper
and M. J. Duff, ``Trace Anomalies In Dimensional Regulariztion,''
Nuovo Cim. {\bf A23} (1974) 173}). Though the one-loop effective
action is non-local and is not Weyl-invariant, the conformal
anomaly is local and Weyl-invariant. Since it is also
supersymmetric, it has the full local superconformal symmetry.
Moreover, given the structure of the one-loop conformal anomaly,
the part of this functional that involves the Weyl tensor is
simply $\int d^4x \sqrt g \,W^2$, rather than $\int d^4x \sqrt g
\,h(C,\bar C)W^2$ for some non-trivial function $h(C,\bar C)$.}

By contrast, it appears that  twistor-string theory corresponds to
%NBnew  sign of W?
$f({\cal W}) = e^{2{\cal W}}$. With  this choice of $f$, we lose
the $SL(2,\Bbb{R})$ symmetry that the classical theory very
plausibly possesses if $f$ is quadratic.  However, it is
conceivable that twistor-string theory has an $SL(2,\Bbb{Z})$
symmetry.   Possibly, such a symmetry acts as $SL(2,\Bbb{Z})$ on
the pair $(W,W_D)$, where $W_D=\partial f/\partial W$; this would
be roughly analogous to what happens in four-dimensional gauge
theories with ${\cal N}=2$ supersymmetry \ref\sw{N. Seiberg and E.
Witten, ``Electric-Magnetic Duality, Monopole Condensation, and
Confinement In ${\cal N}=2$ Supersymmetric Gauge Theory,'' Nucl.
Phys. {\bf B426} (1994) 19, hep-th/9407087.}.  Such a duality
symmetry would entail non-classical electric-magnetic duality
transformations on the metric tensor in spacetime, and in that
respect would differ from presently known dualities in gauge
theory and string theory.

\newsec{Anomalies And Gauge Groups}

In this section, we consider constraints associated with
anomalies.  First we consider implications of anomalies for the
gauge group, and then we analyze some issues involving anomalies
that are special to the $\Bbb{CP}^{3|4}$ approach to
twistor-string theory.

\subsec{Constraints On The Gauge Group}

For physical Type I and heterotic strings, the possible gauge
groups are determined by Green-Schwarz anomaly cancellation, or
alternatively, by cancellation of certain worldsheet tadpoles
and/or anomalies.  What happens in twistor-string theory?

Here, there are two obvious constraints on the gauge group.  They
appear to lead to somewhat different answers, a point that at the
moment we cannot illuminate.

One constraint arises because of the $SU(4)$ $R$-symmetry of
${\cal N}=4$ super Yang-Mills theory.  This symmetry is gauged
when the super Yang-Mills theory is coupled to conformal
supergravity; indeed, the $SU(4)_R$ gauge fields, with helicities
$\pm 1$ and transforming in the ${\bf 15}$ of $SU(4)_R$, are part
of the spectrum that we analyzed in sections 2 and 3.

The $SU(4)_R$ symmetry is potentially anomalous.  For example, in
the vector multiplet, the massless helicity $1/2$ fields transform
as ${\bf {\bar 4}}$ while the helicity $-1/2$ fields transform as
the ${\bf 4}$.  So the vector multiplets give an $SU(4)_R^3$
anomaly with coefficient $-{\rm dim}\,G$, with ${\rm dim}\,G$ the
dimension of the gauge group $G$.  By contrast, as analyzed in
\ref\romer{H. Romer and P. van Nieuwenhuysen, ``Axial Anomalies In
${\cal N}=4$ Conformal Supergravity,'' Phys. Lett. {\bf 162B}
(1985) 290.}, the conformal supergravity multiplet has anomaly
$+4$.\foot{Our convention for $SU(4)$ quantum numbers of fields
and hence for the sign of the anomaly is opposite to that in
\romer.} The authors of \romer\ threfore conclude that an
anomaly-free theory must have ${\rm dim}\,G=4$, so $G=SU(2)\times
U(1)$ or $U(1)^4$.

On the other hand, we can approach the matter using the worldsheet
conformal anomaly.  In the open-string approach to twistor-string
theory, gauge symmetry arises when the matter system (whose
Lagrangian is called $S_C$ in \actions) includes a current
algebra. The only known general restriction on this matter system
is that \berk\ it must have $c=28$. This permits a wide variety of
current algebras, and certainly does not determine what the
dimension or rank of the gauge group must be.  $SU(2)\times U(1)$
and $U(1)^4$ are possible, but are not forced upon us.\foot{One
exotic possibility is that the current algebra is described by the
product of $\Bbb{T}^4$ and the $c=24$ ``Monstrous moonshine''
conformal field theory of \ref\moon{I.B. Frenkel, J. Lepowski and
A. Meurman, in ``Vertex Operators in Mathematics and Physics,''
Publications of the Mathematical Sciences Research Institute No. 3
(Springer, Berlin, 1984).}. Since the monstrous moonshine
conformal field theory has no dimension one fields, the only
contribution to the spectrum would come from the four free bosons
of $\Bbb{T}^4$ which could be used to construct vertex operators
for $U(1)^4$ gauge fields.} Conceivably, from an open-string point
of view, the $c=28$ constraint must be supplemented by additional,
presently unknown, restrictions involving tadpole cancellation in
perturbation theory.

\nref\frad{E.S. Fradkin and A.A. Tseytlin, ``Conformal Anomaly in
Weyl Theory and Anomaly Free Superconformal Theories,'' Phys.
Lett. {\bf B134} (1984) 187.}%

 Apart from the $SU(4)_R$ anomaly,
one should consider other potential anomalies such as the
conformal anomaly \refs{\confgrav,\frad}.  We would presume, but
have not demonstrated, that these anomalies are all related by
${\cal N}=4$ supersymmetry, and vanish when the $SU(4)_R^3$
anomaly does. If this is so, then as the $SU(4)_R^3$ anomaly is a
chiral anomaly that arises only in one-loop order, cancellation of
this anomaly presumably entails cancellation of the conformal
anomalies and other anomalies to all orders.

\bigskip\noindent{\it The Level Of The Current Algebra}

Requiring $c=28$ does not determine the symmetry group $G$ of the
current algebra, and likewise does not determine the {\it level}
of the current algebra, which we will call $k$. For $SU(2)$, a
current algebra at level $k$ has $c_{SU(2)}=3k/(k+2)$, so there is
``room'' for many values of $k$, even if we want the $c=28$ system
to be unitary.

In the discussion of scattering amplitudes in section 5, $k$
really only enters because the connected part of the correlation
function \eqn\hugul{\langle J^{r_1}(z_1)\dots J^{r_s}(z_s)\rangle}
is proportional to $k$.  Because of this, just as for the
heterotic string, the Yang-Mills effective action in spacetime is
proportional to $k$. (The ten-dimensional heterotic string has
$k=1$, but compactified models can be constructed with various
values of $k$.)  The kinetic energy for gauge and gravitational
fields is thus qualitatively \eqn\plxko{{1\over g_s^2}\int
d^4x\left(k\Tr\,F^2+W^2\right),} with $g_s$ the string coupling
constant, $F$ the Yang-Mills field strength, and $W$ the Weyl
curvature. As explained in section 5, $g_s$ is best understood as
arising from the expectation value of the dilaton field, but this
is not important at the moment. From \plxko, we see that to
decouple conformal supergravity, we should take the limit $k\to 0$
with $k/g_s^2$ fixed.

Another way to explain why conformal supergravity decouples for
$k\to 0$ is to note the following.  We express the argument for
the case $G=U(N)$.  In $U(N)$ current algebra in genus zero, the
single-trace part of the current correlators \hugul\ are
proportional to $k$, while multi-trace contributions are
proportional to higher powers of $k$.  So for $k\to 0$, if we
adjust $g_s^2$ to cancel one power of $k$, the correlation
functions reduce to single-trace expressions.  The single-trace
part of the genus zero correlation functions reproduce Yang-Mills
scattering amplitudes, while (as explained in section 5.1 of
\witten) multi-trace contributions to correlation functions
reflect contributions from exchange of supergravitons.

Unitarity of the current algebra requires $k$ to be a positive
integer, and even if one does not care about unitarity, for the
current algebra to be defined globally normally requires that $k$
should be an integer.  So at the moment it is difficult to see how
to make sense in the string theory of the limit $k\to 0$, $k/g_s^2$
fixed.

\subsec{Anomalies In The $B$-Model Of $\Bbb{CP}^{3|4}$}

Now we briefly consider some issues involving anomalies in the
other approach to twistor-string theory.

The first question we ask is this: what is the analog for the
$B$-model of $\Bbb{CP}^{3|4}$ of the constraint $c=28$?\foot{These
issues have also been investigated by M. Movshev (private
communication) with similar conclusions.}

In this $B$-model, scattering amplitudes are computed by
integration over a suitable moduli space ${\cal M}$ of curves
$\Bbb{D}\subset \Bbb{CP}^{3|4}$.  ${\cal M}$ is a complex
manifold.  To integrate over ${\cal M}$, one needs to find a
suitable holomorphic measure $\Theta$ on ${\cal M}$.  Then the
integration is defined as $\int_{\cal M_{\Bbb{R}}}\Theta$, where
${\cal M}_{\Bbb{R}}$ is a suitable real cycle in ${\cal M}$.

We consider the case of $N$ $D5$-branes on $\Bbb{CP}^{3|4}$,
corresponding to gauge group $U(N)$. We have to integrate over the
space of triples consisting of

{\it (i)} a Riemann surface $\Bbb{D}$;

{\it (ii)} a holomorphic map $\Phi:\Bbb{D}\to \Bbb{CP}^{3|4}$;

{\it (iii)} and worldsheet fermions $\alpha^i$, $\beta_j$,
$i,j=1,\dots,N$ transforming under $U(N)$  as ${\bf N}$ and ${\bf
{\bar N}}$.

Suppose that we were merely trying to integrate over the moduli
space ${\cal M}_0$ of abstract Riemann surfaces $\Bbb{D}$. There
is no natural holomorphic measure, roughly speaking since ${\cal
M}_0$ is not a Calabi-Yau manifold.  A holomorphic measure would
be a holomorphic section of the canonical bundle $K_{{\cal M}_0}$
of ${\cal M}_0$, but this line bundle is non-trivial.  Instead,
holomorphic factorization of the bosonic string \ref\knizmore{A.
A. Belavin and V. G. Knizhnik, ``Algebraic Geometry And The
Geometry Of Quantum Strings,'' Phys. Lett. {\bf B168} (1986) 201;
V. G. Knizhnik, ``Multiloop Amplitudes In The Theory Of Quantum
Strings And Complex Geometry,'' Sov. Phys. Usp. {\bf 32} (1989)
945.}  is based on the isomorphism $K_{{\cal M}_0}\cong
\lambda_0^{13}$, where $\lambda_0$ is the determinant line bundle
of the $\bar\partial$ operator acting on ordinary functions.
Existence of this isomorphism is equivalent to the statement that
\eqn\ucucc{K_{{\cal M}_0}\otimes \lambda_0^{-13}} is trivial and
so has a natural holomorphic section.  The measure of the $b-c$
ghost system of the bosonic string is a section of $K_{{\cal
M}_0}$. The chiral part of the measure of a complex boson (or two
real bosons) is a section of $\lambda_0^{-13}$.  So by including
13 complex bosons, or more generally a chiral matter system with
$c=26$, we do obtain a holomorphic measure.

If the integration data were purely the Riemann surface $\Bbb{D}$
and the $\alpha-\beta$ system, that is {\it (i)} and {\it (iii)}
above, then we would infer from this that the $\alpha-\beta$
system should have $c=26$.  Actually, as we explain momentarily,
part {\it (ii)} of the data, the holomorphic map $\Phi:\Bbb{D}\to
\Bbb{CP}^{3|4}$, effectively carries $c=-2$, so the $\alpha-\beta$
system should have $c=28$.  As a single pair of fermions
$\alpha,\beta$ has $c=1$, this seems to mean that we should take
$N=28$, leading to $U(28)$ gauge theory at level one coupled to
conformal supergravity. Of course, from the standpoint of the
open-string approach to twistor-string theory, $U(28)$ at level
one is just one of many possibilities.

Given an abstract Riemann surface $\Bbb{D}$, we will now examine
the holomorphic maps $\Phi:\Bbb{D}\to \Bbb{CP}^{3|4}$ of
sufficiently high degree $d$.  (For maps of low degree, the
conclusion we will reach is still valid, but a more detailed
argument is needed.)  The key is to introduce the line bundle
${\cal O}(1)$ over $\Bbb{CP}^{3|4}$, and define a line bundle
${\cal L}$ over $\Bbb{D}$ as the pullback ${\cal L}=\Phi^*({\cal
O}(1))$. For sufficiently high $d$, ${\cal L}$ can be any line
bundle over $\Bbb{D}$ of degree $d$. Once ${\cal L}$ is picked,
the map $\Phi$ is determined by picking four bosonic holomorphic
sections $(s^a,s^{\dot a})$ of ${\cal L}$ and four fermionic ones
$s^A$; $\Phi$ is then defined by setting $Z^I=s^I$.  The possible
$\Phi$'s for given ${\cal L}$ are parametrized by the choices of
$s^I$ modulo an overall scaling $s^I\to ts^I$, $t\in \Bbb{C}^*$.
The space of $s^I$ modulo this scaling is a copy of
$\Bbb{CP}^{M-1|M}$ (for some $M$) and has a natural holomorphic
measure, simply because the bosonic and fermionic variables in
$s^I$ are equal in number and have the same quantum numbers. (This
statement is explained in more detail in section 4.6 of \witten,
where it is used to construct the integration measure for the case
that $\Bbb{D}$ has genus zero.) So choosing a measure for
integrating over maps $\Phi:\Bbb{D}\to\Bbb{CP}^{3|4}$ for fixed
$\Bbb{D}$ amounts to choosing a measure for the space of line
bundles ${\cal L}$.

The tangent space to the space of ${\cal L}$'s is
$H^1(\Bbb{D},{\cal O})$, so a measure on the space of ${\cal L}$'s
is a section of the line  bundle (over ${\cal M}_0$) whose fiber
is $H^1(\Bbb{D},{\cal O})^{-1}$.  This is the bundle which above
was called $\lambda_0$.  Since the measure on the variables of
type ({\it ii}) is thus effectively a holomorphic section of
$\lambda_0$, to use the triviality of $K_{{\cal M}_0}\otimes
\lambda_0^{-13}$ in constructing a measure for the overall system,
the integration measure for the ``matter'' system must be a
section of $\lambda_0^{-14}$, that is, the matter system must have
$c=28$.

The choice here of ${\cal L}$  is related in the open string
analysis to the $GL(1)$ scaling, reviewed in section 2, which
similarly (according to \berk) shifts the central charge required
for the matter system from $c=26$ to $c=28$.  The $D1$-brane in
$\Bbb{CP}^{3|4}$ also carries a $U(1)$ gauge field (from the
$D1-D1$ strings) that has no obvious counterpart in the
open-string approach to twistor-strings, and which we have
neglected above.  Its role really merits further study.

\bigskip\noindent{\it A Holomorphic Green-Schwarz Mechanism}

For the heterotic string, at least in the context of
compactification, there are really two stages to worldsheet
anomaly cancellation.  Via the Green-Schwarz mechanism, worldsheet
gauge and gravitational anomalies are canceled.  Once this is
done, it makes sense to discuss the $c$-number conformal anomaly.

Both of these issues have analogs for the $B$-model of
$\Bbb{CP}^{3|4}$.  We have already explored the analog of the
conformal anomaly.  Now let us briefly discuss the analogs of
worldsheet gauge and gravitational anomalies.

The fermions $\alpha$ and $\beta$ on the $D$-instanton worldvolume
$\Bbb{D}$ have an action $\int_{\Bbb{D}}\beta
\bar\partial_A\alpha$, where $\bar\partial_A=\bar\partial+A$ is
the chiral Dirac operator acting on $\alpha$. Here $A$ is the
$U(N)$ gauge field on $\Bbb{CP}^{3|4}$, or rather its restriction
or pullback to $\Bbb{D}$.  Accordingly, the path integral over
$\alpha$ and $\beta$ gives a factor $\det\,\bar\partial_A$.  Since
$\bar\partial_A$ is a chiral Dirac operator, this determinant has
a gauge anomaly.  Under a gauge transformation $\delta
A=\bar\partial_A\epsilon$, with $\epsilon$ an infinitesimal gauge
parameter, we have
\eqn\sononon{\det\,\bar\partial_A\to\exp\left({1\over
2\pi}\int_{\Bbb{D}}d\bar z\wedge dz\,\Tr\,A_{\bar
z}\partial_z\epsilon\right) \,\det\,\bar\partial_A.} Here $z$ is a
local holomorphic coordinate on $\Bbb{D}$.  So gauge invariance
appears to be lost.

What saves the day, just as for the heterotic string, is the
coupling to the $B$-field.  The worldsheet path integral is more
accurately represented as
\eqn\ononon{\exp\left(-\int_{\Bbb{D}}b_{z\bar z}dz\wedge d\bar
z\right)\,\det\,\bar\partial_A,} so all is well if under gauge
transformations \eqn\hnno{b_{I\bar J}\to b_{I\bar J}+{1\over
2\pi}\Tr\,A_{\bar J}\partial_I\epsilon.}

\nref\wwitten{E. Witten, ``Chern-Simons Gauge Theory As A String
Theory,'' Prog. Math. {\bf 133} (1995) 637, hep-th/9209074.}%
This gauge transformation law for $b$ can be deduced on other
grounds. Let us consider the Chern-Simons $(0,3)$-form action for
$A$ \refs{\wwitten,\witten},\foot{We will not be precise with our
coefficients, so from this point of view, we will not verify the
coefficient in \hnno.} \eqn\unon{\int d\bar X^{\bar I}\wedge d\bar
X^{\bar J}\wedge d\bar X^{\bar K}\Tr\left(A_{\bar I}\partial_{\bar
J}A_{\bar K}+{2\over 3}A_{\bar I}A_{\bar J}A_{\bar
K}\right)\Omega.}  (Local coordinates $X^I$ are used here as in
\yturu.)  This action can be defined for any almost complex
structure $J$.  In verifying that it is gauge invariant, one uses
$\bar\partial^2=0$, which only holds for integrable complex
structures.  In general, if the Nijenhuis tensor $N(J)$ is
nonzero, the gauge variation of \unon\ is  \eqn\turmoil{\int d\bar
X^{\bar I}d\bar X^{\bar J}d\bar X^{\bar K}\Tr\,\left(A_{\bar
I}\partial_L\epsilon\right)N_{\bar J\,\bar K}^L\,\Omega.}

In our discussion of the closed string modes in section 2.2, we
did not take $N(J)=0$ by definition.  Rather, $N(J)=0$ is the
equation of motion for the field $b$, derived from the action
\yturu.  If we postulate that $b$ transforms under gauge
transformations as in \hnno\ (and adjust a couple of
coefficients), the sum of \yturu\ and \unon\ becomes
gauge-invariant.  Thus, the gauge transformation law of $b$ that
restores gauge-invariance in the $D$-instanton path integral is
also needed to ensure gauge-invariance of the bulk effective
action on $\Bbb{CP}^{3|4}$.

There is apparently a similar story for diffeomorphism anomalies.
The Chern-Simons action $\int \omega_{CS}(J)\Omega$ of eqn.
\nxxno\ is diffeomorphism invariant only if $N(J)=0$.  But the sum
of \yturu\ and \nxxno\ is diffeomorphism-invariant if one adds a
gravitational contribution to the transformation law \hnno\ of
$b$.  This contribution is analogous to the transformation of the
$B$-field of the heterotic string under diffeomorphisms, and also
serves to cancel the gravitational anomaly in the $D$-instanton
measure.

\vskip 15pt

{\bf Acknowledgements:} We would like to thank Freddy Cachazo,
Lubos Motl, Ilya Shapiro, Warren Siegel, Peter Svrcek, Arkady
Tseytlin, Cumrun Vafa and Brenno Carlini Vallilo for useful
discussions. NB would also like to acknowledge partial financial
support from CNPq grant 300256/94-9, Pronex grant 66.2002/1998-9,
and FAPESP grant 99/12763-0, and to thank the Institute for
Advanced Study and the Funda\c{c}\~ao do Instituto de F\'{\i}sica
Te\'orica for their hospitality.  EW acknowledges the support of
NSF Grant PHY-0070928.

\listrefs

\end